\documentclass[%
 reprint,
 amsmath,amssymb,
aps,
prb,
floatfix
]{revtex4-1}

\usepackage{graphicx}
\usepackage{dcolumn}
\usepackage{bm}

\usepackage{nicefrac}
\usepackage[usenames, dvipsnames]{xcolor}
\usepackage{subfigure}
\usepackage{threeparttable}

\begin{document}


\title{Hidden scale invariance at high pressures in gold and five other fcc metal crystals}

\author{Laura Friedeheim}
\email{lauraf@ruc.dk}
\author{Jeppe C. Dyre}
\author{Nicholas P. Bailey}%


\affiliation{
 ``Glass and Time'', IMFUFA, Department of Science, Roskilde University, P.O. Box 260, DK-4000 Roskilde, Denmark
}%


\date{\today}

\begin{abstract}

Recent DFT (density functional theory) simulations showed that metals have a hitherto overlooked symmetry termed ``hidden scale invariance'' [Hummel {\em et al.}, Phys. Rev. B {\bf{92}}, 174116 (2015)]. According to isomorph theory, this scaling property implies the existence of lines in the thermodynamic phase diagram, so-called isomorphs, along which structure and dynamics are invariant to a good approximation when given in properly reduced units. This means that the phase diagram becomes effectively one-dimensional with regard to several physical properties. 
This paper investigates consequences and implications of the isomorph theory in six metallic crystals; Au, Ni, Cu, Pd, Ag and Pt. The data are obtained from molecular dynamics simulations employing many body 'effective medium theory' (EMT) to model the atomic interactions realistically. We test the predictions from isomorph theory for structure and dynamics by means of the radial distribution and the velocity autocorrelation functions, as well as the rather dramatic prediction of instantaneous equilibration after a jump between two isomorphic points. Many properties of crystals tend to be dominated by defects and many of the properties associated with these defects are expected to be isomorph invariant as well. This is investigated in this paper for the case of vacancy diffusion. We find the predicted invariance of structure and also of dynamics, though less rigorous. We show results on the variation of the density scaling exponent $\gamma$, which can be related to the Gr\"uneisen-parameter, for all six metals. We consider large density changes up to a factor of two, corresponding to very high pressures. Unlike systems modelled using the Lennard-Jones potential where the density scaling-exponent $\gamma$ is almost constant, it varies substantially when using the EMT potential and is also strongly material dependent. 
\end{abstract}

\maketitle


\newcommand{\Sex}{S_{\textrm{ex}}}
\newcommand{\R}{\boldsymbol{R}}
\newcommand{\fig}[1]{Figure~\ref{#1}}

\section{Introduction}

The most common state of metals as used by humans is the solid (crystal) phase. Investigation of the properties of pure crystalline metals has played a huge role in the development of solid state physics \cite{Ashcroft/Mermin:1976}, and the mechanical properties of pure metals and alloys have historically been the most important topic in materials science \cite{Gottstein:2004}. It might therefore be thought that all of the basic physics of pure crystalline metals have been well understood and documented. However, recent work has demonstrated the existence of a kind of previously unknown scale invariance in a range of model systems, including metals, in both the liquid and crystal phases. Specifically, in the part of the phase diagram corresponding to the condensed phases there exist curves, termed {\it isomorphs}, along which a large set of physical properties, namely those relating to structure and microscopic dynamics, as well as some thermodynamic properties and some transport coefficients, are approximately invariant when expressed in appropriately scaled units \cite{paper4}. Recent {\it ab initio} simulations \cite{Hummel/others:2015} have confirmed that many pure metals belong to the class of systems which have good isomorphs, a class known as Roskilde or R-simple systems. It is the purpose of this paper to document isomorph invariance of structure and dynamics of perfect metallic crystals, specifically the fcc metals Au, Ni, Cu, Pd, Ag, Pt. The work was inspired by a bachelor student project which investigated isomorphs in the liquid state for the same six metals\cite{bachelorthesis:2017}.

An early paper \cite{paper1} presented some evidence that metallic systems belong to the class of R-simple systems. Hu et al. have also reported results for a simulated metallic glass \cite{Hu/others:2016}. Recently, Hummel {\it et al.} confirmed using density functional theory (DFT) methods that most metals are R-simple close to their triple point \cite{Hummel/others:2015}. Because of the large computational cost of DFT methods, other state points were not studied, so the variation of for example the density scaling exponent $\gamma$ has not been studied. Moreover, the cost of DFT calculations limits what aspects of thermodynamics and structure can be studied, and essentially prohibits the study of dynamics and transport coefficients. It is of great interest to investigate and document expected isomorph variances in metallic crystals, liquids, and amorphous structures (metallic glasses) using many-body empirical potentials, which offer a reasonable compromise between computational efficiency and accuracy. In addition, metals form an interesting class of R-simple systems because they are not described by pair interactions (as evidenced by the violation of the Cauchy relations for the elastic constants) \cite{wallace:1998thermodynamics}; while a good understanding of the density-scaling properties of systems with pair interactions exists \cite{Andersen/Weeks/Chandler:1971, Roland/Bair/Casalini:2006, Ingebrigtsen/others:2012}, many-body systems present a challenge: are they R-simple? 

In this work we use the effective medium theory (EMT) semi-empirical many-body potential \cite{Jacobsen/Stoltze/Norskov:1996}. It is considered semi-empirical because it is derived from DFT, and some of the parameters are drawn directly from DFT calculations. The  expression for the total potential energy is similar in structure to other commonly used many-body potentials for metals, such as the embedded atom method (EAM), involving pair-sums and some nonlinear ``embedding'' function. Unlike many EAM potentials, EMT is based on fairly simple functional forms, rather than complex functions which require heavy fitting to large data sets and are typically tabulated. This means that 
(1) EMT has been relatively straightforward to implement in our graphical processing unit (GPU) molecular dynamics software RUMD \cite{Bailey/others:2017, RUMD} and 
(2) we can hope to understand analytically the existence of strong virial potential-energy correlation in this potential and moreover find an analytic expression for how $\gamma$ depends on density. We use the simplest version of EMT presented in Ref.~\onlinecite{Jacobsen/Stoltze/Norskov:1996}, which provides all parameters necessary to simulate EMT models of Ni, Cu, Pd, Ag, Pt and Au.

We restrict our investigation of the isomorph scaling properties of metallic systems to the crystal phases of pure systems, the metal elements listed above. These all have a face centred cubic (fcc) ground state at zero pressure. A previous work considered the isomorph scaling properties of classical crystals consisting both of spherical particles interacting via pair potentials, as well as simple molecular systems, and found that simple measures of structure and dynamics are invariant along isomorphs, as expected \cite{Albrechtsen2014}. We consider the same properties as those authors: we investigate structure as quantified by the radial distribution function (RDF) and dynamics as quantified by the velocity autocorrelation function (VAF), which can be related to the phonon spectrum \cite{Changyol/others:1993}. Mechanical properties of crystalline materials tend to be dominated by defects, specifically vacancies, interstitials, dislocations, stacking faults and grain boundaries \cite{kelly/knowles:2012}. Many properties associated with defects are expected to be isomorph invariant - for example defect mobilities - when expressed in reduced units. As in Ref.~\onlinecite{Albrechtsen2014} we investigate in this work a simple case, namely vacancy diffusion. We also check one of the dramatic predictions of isomorph theory, instantaneous equilibration when a system is brought rapidly from one point to another on the same isomorph \cite{paper4}. 

\section{Isomorph theory and  hidden scale invariance}



Isomorph theory has been developed throughout a series of papers \cite{paper1,paper2,paper3,paper4,paper5} starting from first establishing the existence and subsequently developing a theoretical understanding of strong correlations between the equilibrium fluctuations of the configurational parts of pressure and energy. The correlations are deemed strong when $R > 0.9$ where $R$ is the (Pearson) correlation coefficient
\begin{equation}
    R=\frac{\langle \Delta W \Delta U \rangle}{\sqrt{\langle (\Delta W)^2\rangle \langle(\Delta U)^2\rangle}}
    \label{eq:R}
\end{equation}
with the sharp brackets denoting the canonical constant-volume (NVT) averages and where $W$ and $U$ are the virial and the potential energy, respectively. 
Systems with these strong correlations are also referred to as R-simple systems to a) account for the ambiguity of the term `strongly correlated' in physics and chemistry, to b) stress the fact that these systems exhibit a particularly simple behaviour in terms of structure and dynamics, and c) that this behaviour is not limited to liquids only but extends to the solid phase as well since the strong correlations generally appear when the system is dense \cite{paper1, paper2}.  %

%
Paper IV \cite{paper4} of the series mentioned above introduced the concept of \textit{isomorphs}. Isomorphs are curves in the  phase diagram along which certain static, dynamic and thermodynamic quantities are invariant when given in appropriately reduced units. %
%
%
Any configuration can be described in terms of the particle coordinates as 

\begin{equation}
    \boldsymbol{R} = (\boldsymbol{\vec{r}_1}, \boldsymbol{\vec{r}_2},... \boldsymbol{\vec{r}_N})
\end{equation} 
where $\boldsymbol{\vec{r}_i}$ is the coordinate vector of the i-th particle. The reduced unit version is given by $\boldsymbol{\tilde{R}} = \rho^{1/3} \boldsymbol{R}$. If two configurations from different state points have the same reduced coordinates,

\begin{equation}
    \rho_1^{1/3} \boldsymbol{R}_1 = \rho_2^{1/3} \boldsymbol{R}_2.
\end{equation} 
then Roskilde simplicity implies they have approximately proportional configurational NVT Boltzmann factors

\begin{equation}\label{proportional_Boltzmann}
    \exp \left( - \frac{U ( \boldsymbol{R}_1 )}{ k_B T_1} \right) \cong C_{12} \exp \left( - \frac{U ( \boldsymbol{R}_2 )}{ k_B T_2} \right),
\end{equation}
where the constant $C_{12}$ depends only on the state points $(T_1, \rho_1)$ and $(T_2, \rho_2)$ and not on the configurations. This means that the potential energy of a given configuration $U(\boldsymbol{R_i})$ and density $\rho_i$ can be scaled to any configuration on the same isomorph as follows:

\begin{equation}\label{proportional_scaled_energies}
    U ( \boldsymbol{R}_2 ) \cong \frac{T_2}{T_1}U ( \boldsymbol{R}_1 ) + k_B T_2 C_{12} ,
  \end{equation}
or, considering fluctuations about the respective mean values:

\begin{equation}\label{proportional_energy_fluctuations}
   \Delta U ( \boldsymbol{R}_2 ) \cong \frac{T_2}{T_1} \Delta U ( \boldsymbol{R}_1 ).
  \end{equation}

The shape of an isomorph is characterized in terms of the scaling-exponent $\gamma$ defined as the logarithmic derivative of temperature with respect to density along a curve of constant excess entropy. Statistical mechanics provides an expression in terms of fluctuations for this derivative \cite{paper4}, giving:
\begin{equation}
    \gamma 
    \equiv \left( \frac{\partial \ln T}{\partial \ln \rho}  \right)_{\Sex} = \frac{\langle \Delta W \Delta U \rangle}{\langle (\Delta U)^2 \rangle}.
    \label{eq:gamma}
\end{equation}

\begin{figure}[t!]
    \includegraphics[width=8.66cm, draft=false]{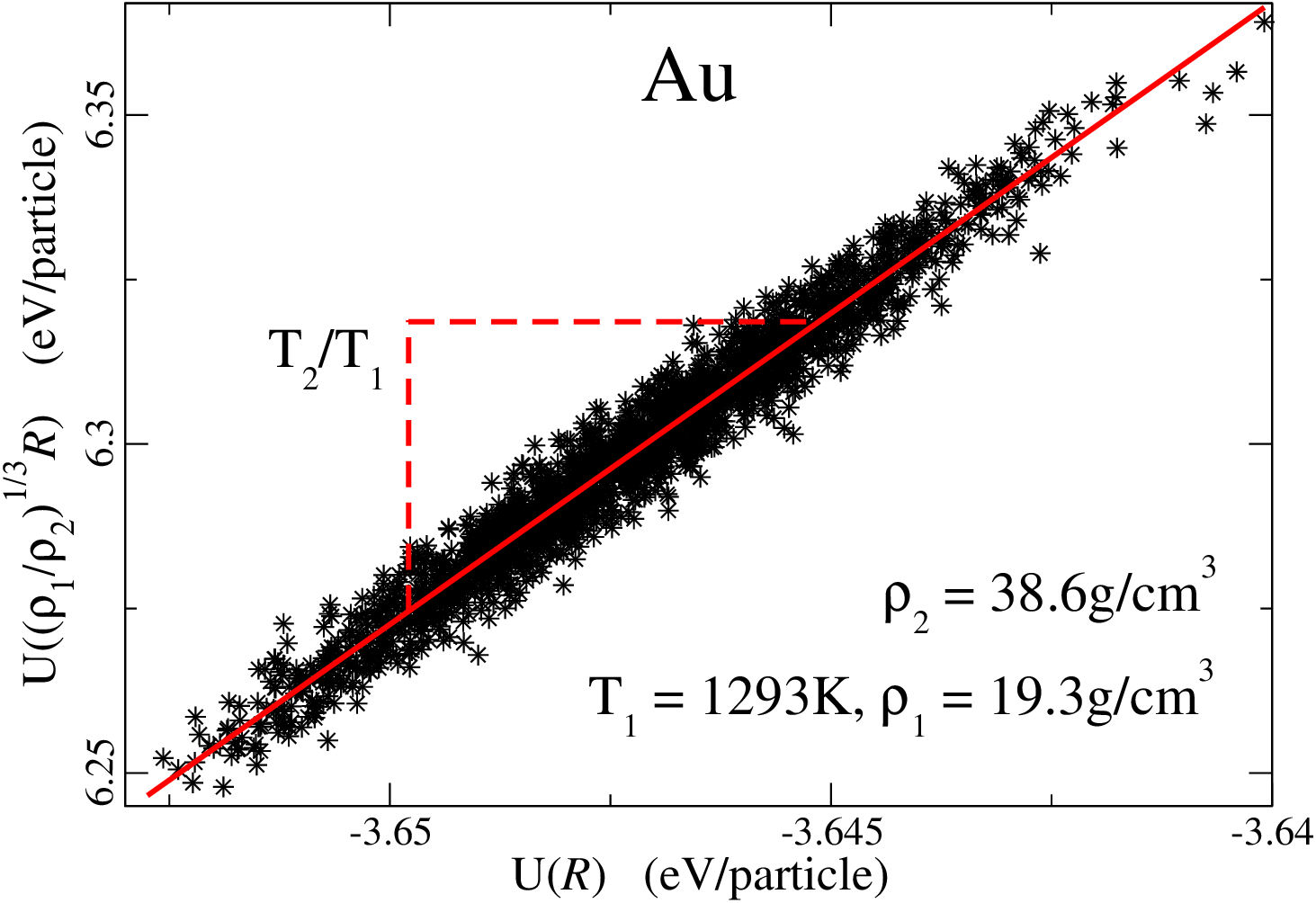}
    \caption{The direct isomorph check for gold: a scatter plot of potential energies of  configurations drawn from a simulation at a given density $\rho_1$ and temperature $T_1$ versus the potential energies of the configurations scaled to another density $\rho_2$. 
    The red line is the best fit line and has the slope $T_2/T_1$, so the temperature $T_2$ for a state point with density $\rho_2$ on the same isomorph as the initial state point can be identified from linear regression. The same initial configuration can be scaled to different densities, thus allowing to map out several isomorphic points from just one simulation. 
    }
    \label{DIC}
\end{figure}

`Excess' quantities are defined in reference to the respective quantity for the ideal gas at the same temperature and density, e.g. $\Sex = S - S_{\textrm{id}}$. 
Equation~\eqref{eq:gamma} allows one to map out isomorphs in a step-wise manner by evaluating $\gamma$ at each state point. Another way to trace isomorphs is the so-called direct isomorph check (DIC), which exploits the connection between the energies and temperatures of two state points of Eqs.~\eqref{proportional_scaled_energies} and \eqref{proportional_energy_fluctuations}. Hence, plotting the potential energies of the initial microscopic configurations at $\rho_1$ versus the potential energies of the configurations scaled to another density $\rho_2$ results in a scatter plot where the slope of the best fit line is given by the ratio of the temperatures, $T_2/T_1$. 
An example of such a scatter plot is shown in Figure \ref{DIC}. The advantage of the direct isomorph check is that this method allows a whole isomorph to be generated from only one simulation at one reference point. We have checked that the generated temperatures differ by at most $0.5 \%$  from isomorphs generated in the step-wise manner by Eq.~\eqref{eq:gamma}.

The existence of isomorphs yields the profound simplification of effectively reducing the $(T,\rho)$-phase diagram by one dimension. The one-to-one correspondence between state points as illustrated above also explains why so many quantities are invariant along isomorphs when given in reduced units. 
Using the length unit $l_0$, time unit $t_0$ and an energy unit $e_0$ defined as follows:
\begin{equation}
    l_0 = \rho^{-1/3}, \quad t_0 = \rho^{-1/3}\sqrt{m/k_B T}, \quad e_0 = k_B T,
    \label{eq:redUnits}
\end{equation}
all quantities can be expressed in a dimensionless form to compensate for the trivial scaling of lengths by average interparticle spacing and energies by the temperature.



\begin{figure}[t!]
    \includegraphics[width=8.66cm]{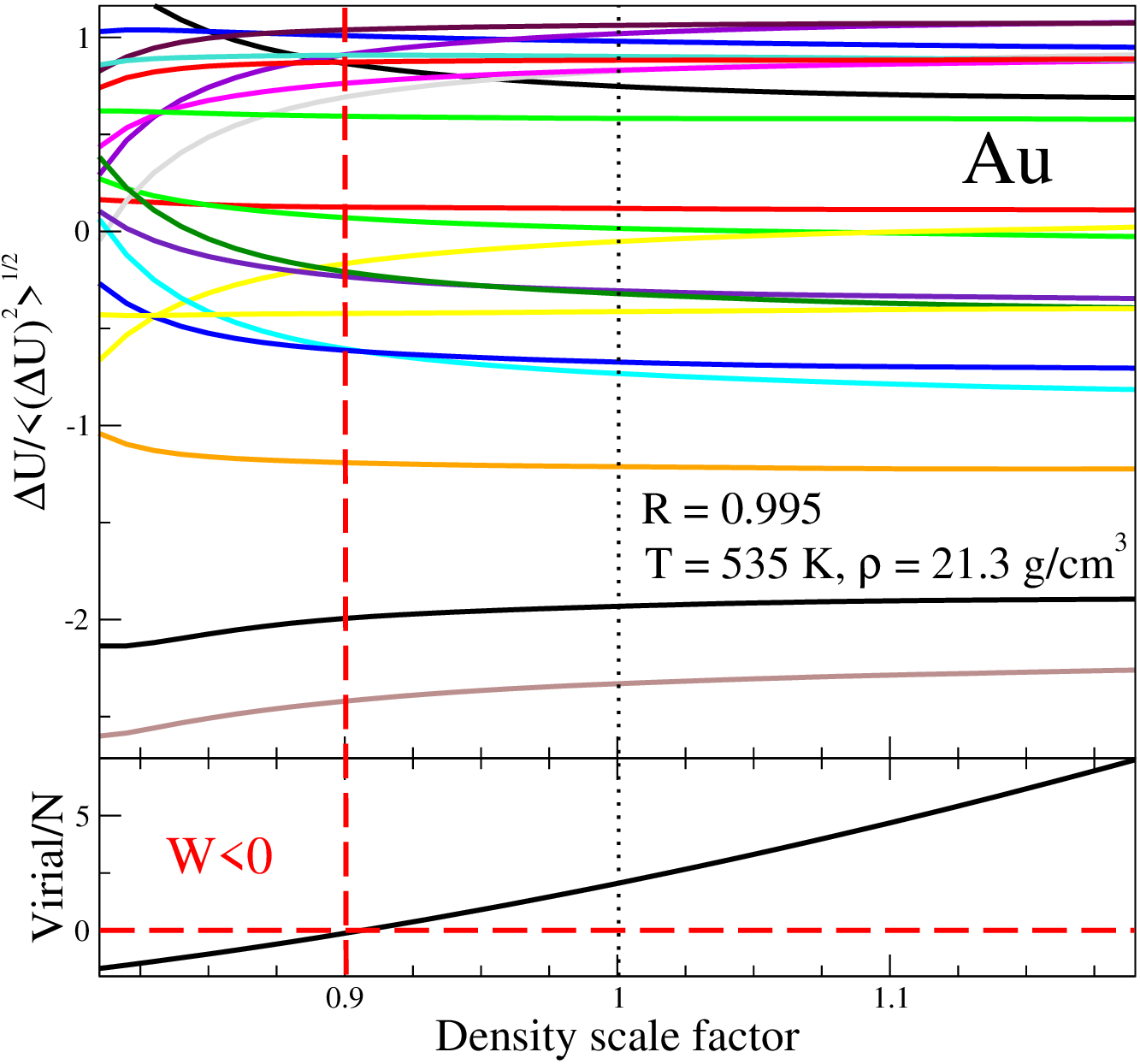}
    \caption{Gold's potential energy per particle after subtracting the average and scaling by the standard deviation of 20 configurations taken from an equilibrium simulation which were subsequently scaled uniformly by 20\% up and down in density and plotted as a function of the density scaling factor. The black line indicates the unscaled energies. The lines obtained in this way illustrate the hidden scale invariance of R-simple systems and cannot cross each other in the ideal ($R=1$) case. %
    The configurations used for this figure are taken from equilibrium simulations at a state point close to ambient conditions and $R=0.995$. The bottom panel shows the virial for the scaled configurations. The strongly diverging lines in the left part of the figure are due to the virial becoming negative (indicated by the red dashed lines).}
    \label{ScalePlot}
\end{figure}

As pointed out in paper IV (\cite{paper4}), systems with strong correlations have isomorphs and vice versa, i.e., these two features are equivalent. It was found later that they are both manifestations of an underlying \emph{hidden scale invariance}. Indeed, isomorph theory has been refined in Ref.~\onlinecite{Schroder/Dyre:2014} by defining R-simple systems directly from their scale invariance. 
It is based on the following simple scaling behaviour

\begin{equation}
    U(\R_a) < U(\R_b) \, \Rightarrow \, U(\lambda \R_a) < U(\lambda \R_b)
\end{equation}
where $U(\R_i)$ is the potential energy of a configuration $\R_i$ and $\lambda$ is a scaling parameter. Thus, a uniform scaling of configurations does not change the ordering of potential energies. For most systems this scale invariance is approximate and dubbed 'hidden' since it is not obvious from the mathematical expression for the potential. This approximate scaling is illustrated in Fig.~\ref{ScalePlot}, where the potential energies of twenty configurations from an equilibrium simulation have been scaled to different densities. For clarity the energies have been shifted and scaled using the mean value and standard deviation at each density. For perfectly isomorphic systems - with correlation coefficient $R=1$ - the lines cannot cross each other. The red dashed line indicates where the virial becomes negative, which leads to a break down of the scaling properties as seen by the sudden diverging of the lines.


The updated definition preserves that isomorphs are the configurational adiabats of the phase diagram, curves along which structure, dynamics and the excess entropy $\Sex$ are invariant together with the simplification of effectively reducing the phase diagram by one dimension. 
Subtle differences between the versions of isomorph theory emanate from the fact that the original formulation is a first order approximation of the more accurate theory of Ref.~\onlinecite{Schroder/Dyre:2014}. This can be illustrated, for example, using the case of the isochoric heat capacity $C_V$.
If exactly obeyed, Eq.~\eqref{proportional_Boltzmann} implies that $C_V$ is invariant along isomorphs, which is often a good approximation but not exact. The slight variation of $C_V$ along isomorphs can, however, be accommodated using the more recent formulation of isomorph theory, with which Eq.~\eqref{proportional_energy_fluctuations} can be derived without requiring Eq.~\eqref{proportional_Boltzmann} or \eqref{proportional_scaled_energies} \cite{Schroder/Dyre:2014}.






One of the more fundamental consequences of the update concerns the scaling exponent $\gamma$. Initially the scaling exponent $\gamma$ was interpreted as being related to an effective inverse power law exponent, which (assuming it to be constant) yields the form $\rho^\gamma/T=$const for isomorphs, consistent with experimental determinations of isochrones \cite{Tolle/others:1998, Tolle:2001, Casalini/Roland:2004, Tarjus/others:2004, Alba-Simionesco/others:2004, Roland/Bair/Casalini:2006, Casalini/Mohanty/Roland:2006}.  Determination of $\gamma$ from fluctuations in simulations shows variation with state point, however \cite{paper1}. It was shown in Ref.~\onlinecite{Ingebrigtsen/others:2012} that the assumption of constant $C_V$ along isomorphs implies that $\gamma$ can only depend on density, which is a fairly good approximation. The most recent definition of hidden scale invariance allows, however, temperature-dependence of $\gamma$ also to be handled within the theory \cite{Schroder/Dyre:2014}.


We find, in fact, that for metals - at least when using the EMT potential - $\gamma$ does vary significantly, both for a given metal and between metals. Table~\ref{gamma_R_DFT_EMT} shows a comparison of the DFT and EMT values of the parameters $R$ and $\gamma$ for the liquid phase near the triple point. There is reasonable agreement between the $R$ and $\gamma$ values, especially noting that the latter vary quite widely, over a factor of two. From this we can conclude that EMT gives a reasonably accurate description of the thermodynamic scaling properties of these metals. A version of this table appeared in Ref.~\onlinecite{bachelorthesis:2017}.

\begin{table}[t]
  \caption{\label{gamma_R_DFT_EMT} Comparison of correlation coefficient $R$ and density scaling exponent $\gamma$ calculated using DFT and using EMT. A liquid state point near the triple point is used in each case. The DFT values are taken from Ref.\cite{Hummel/others:2015}}
\begin{tabular}{|ll|ll|ll|ll|}
\hline 
     &     &      &     &      &       &       &      \\[-.8em]
Sym & Z & $T\,$(K) & $\rho\,$($\frac{g}{cm^3}$) & $R_{EMT}$ &  $R_{DFT}$ & $\gamma_{EMT}$  &$\gamma_{DFT}$ \\
\hline
Ni & 28 & 2000  & 8.19  & 0.96 & 0.92(0.03) & 3.62(0.01) & 3.5(0.3) \\
Cu & 29 & 1480  & 8.02  & 0.95 & 0.90(0.02) & 4.15(0.02) & 4.1(0.2) \\
Pd & 46 & 1900  & 10.38 & 0.91 & 0.92(0.04) & 6.47(0.03) & 4.9(0.5) \\
Ag & 47 & 1350  & 9.32  & 0.93 & 0.90(0.03) & 5.35(0.02) & 4.8(0.4) \\
Pt & 78 & 2200  & 18.53 & 0.87 & 0.87(0.06) & 7.88(0.05) & 6.0(1.4) \\
Au & 79 & 1470  & 16.69 & 0.88 & 0.86(0.14) & 7.93(0.05) & 7.9(1.6)\\
\hline
\end{tabular}
\end{table}

\section{\label{sec:setup} Simulation results}

The results presented in this paper for the fcc metals Ni, Cu, Pd, Ag, Pt and Au have been obtained from simulations carried out in RUMD \cite{Bailey/others:2017, rumd:2017} using the effective medium theory (EMT) potential. 
The potential is based on a reference system modified with a correction term. The reference system is chosen to to give a close to accurate description while still being a simple, well-known system which can be fitted through some build-in scaling parameter. For metal crystals this is can be achieved with an ideal fcc lattice where that lattice constant serves as the scaling parameter. The correction term accounts for the difference in pair-potential between the real and the reference system. A detailed description of the potential and the respective material specific parameters are given in Ref.~\onlinecite{emt}.



\begin{table}[b]
    \caption{\label{tab:Au_data}Pressure $P$, virial $W$, correlation coefficient $R$ and scaling exponent $\gamma$ values along the isomorph for gold. Corresponding tables for the other five metals can be found in the supplemental material.}
  \centering
    \addtolength{\tabcolsep}{2pt}
    \begin{tabular}{ | c  c | c  c | c  c |}
    \hline 
                &     &      &       &       &      \\[-.9em]
       T (K) & $\rho$ ($\frac{g}{cm^{3}}$) & $P$ (GPa) & $W$($\frac{eV}{\text{particle}}$) & $R$ & $\bm{\gamma}$ \\ \hline
1293 & 19.32 & 10 & 0.87 & 0.985 & 6.45 \\ 
2173 & 21.25 & 30 & 3.04 & 0.993 & 4.64 \\
3134 & 23.18 & 70 & 5.60 & 0.996 & 3.78 \\ 
4160 & 25.12 & 110 & 8.51 & 0.996 & 3.26 \\ 
5235 & 27.05 & 160 & 11.74 & 0.997 & 2.90 \\ 
6346 & 28.98 & 220 & 15.26 & 0.997 & 2.64 \\ 
7487 & 30.91 & 300 & 19.05 & 0.997 & 2.44 \\ 
8651 & 32.84 & 380 & 23.07 & 0.998 & 2.28 \\ 
9829 & 34.78 & 480 & 27.30 & 0.998 & 2.14 \\ 
11016 & 36.71 & 590 & 31.73 & 0.998 & 2.02 \\ 
12204 & 38.64 & 710 & 36.33 & 0.998 & 1.92 \\ \hline
    \end{tabular}
    
\end{table}

\begin{figure}
    \centering
    \includegraphics[width=8.66cm]{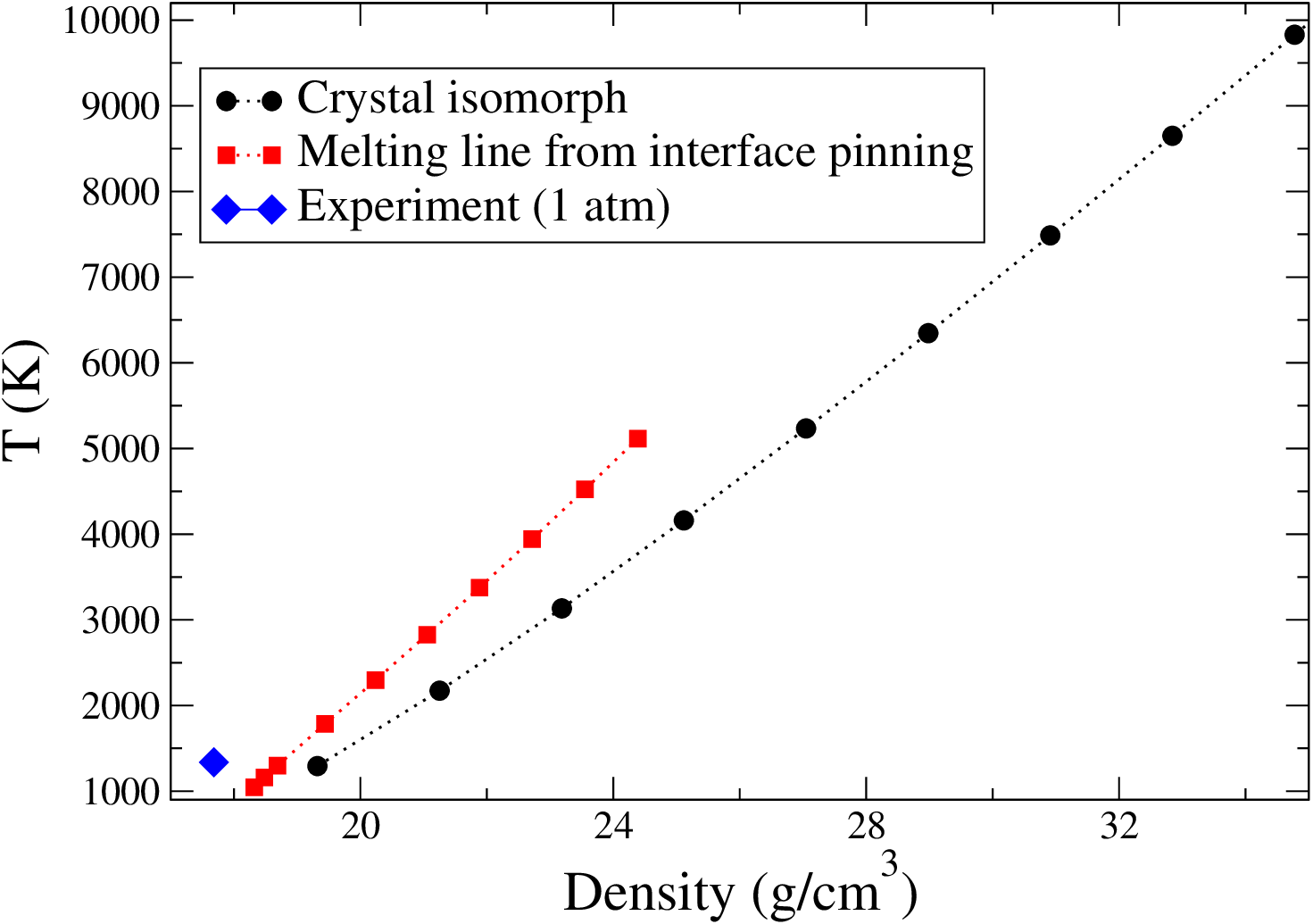}
    \caption{ A density-temperature phase diagram of EMT-Au showing the melting curve  determined by the interface-pinning method \cite{Pedersen:2013} and the crystal isomorph studied in this work. One point from the experimental melting curve is also included \cite{Mirwald/Kennedy:1979}. The isomorphs studied in this work are close to the melting curves.}
    \label{Phase-Diag}
\end{figure}

The simulated systems consist of $4000$ particles organized on a $10 \times 10 \times 10$ fcc lattice with periodic boundary conditions. The initial state point for each metal is chosen to resemble a crystal with room temperature density at $1293$~K ($1000^{\circ}$C) corresponding to a pressure of $10$~GPa. Atomic masses and densities are taken from \cite{Haynes:2014}. For each metal, we simulate three curves: an isomorph, an isotherm and an isochore. The state points for the isomorph have been determined using the direct isomorph check, as described in the previous section, to find isomorphic points corresponding to steps of $10\%$ density change of the reference density up to a total increase of $100\%$ in density. The state points for the isochore (isotherm) are chosen so that they match the temperatures (densities) of the points along the isomorph. 
For each state point the NVT ensemble was simulated using periodic boundary conditions and a Nos\'{e}-Hoover thermostat. Table~\ref{tab:Au_data} shows the temperatures and densities for the isomorph simulated for Au, while Fig.~\ref{Phase-Diag} shows the isomorph together with the melting curve for this system. The latter was determined using the interface pinning method \cite{Pedersen:2013}. A single point from the experimental melting curve is included, showing that it lies somewhat higher in temperature than the model curve.

A slight discrepancy between model and experiment is therefore expected if the model has not explicitly been fitted to the melting temperature. Melting is defined as the point where the Gibbs free energies of the solid and the liquid phase are equal, thus a precise prediction for the melting temperature requires a model that describes both phases with the same accuracy which is usually not the case \cite{Alfe/other:2004}.

\subsection{\label{sec:invariance}Isomorph invariance of structure and dynamics}

\begin{figure}
\centering
    \includegraphics[width=8.66cm]{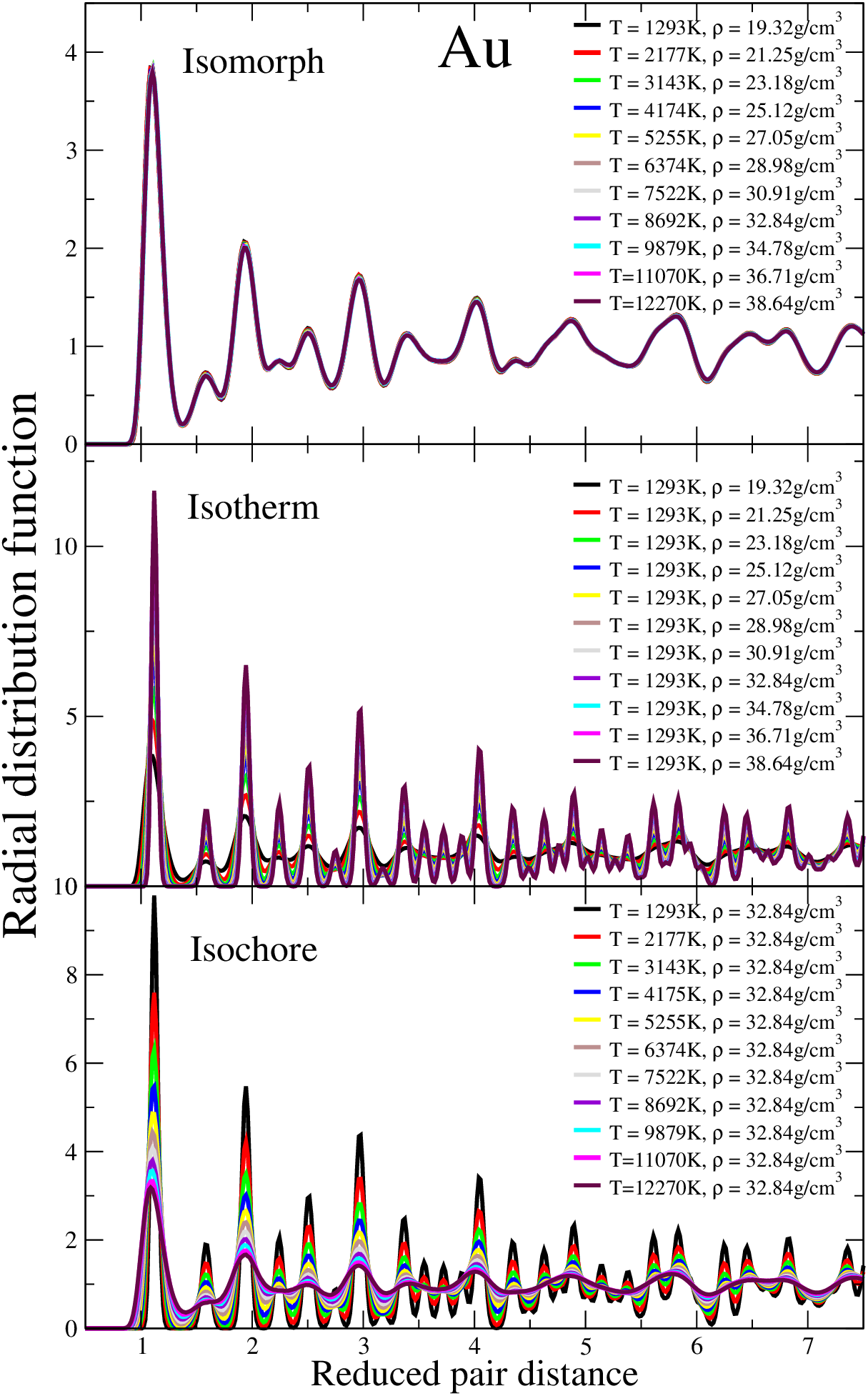}
\caption{The radial distribution functions (RDF) plotted in reduced units (see Eq.~\ref{eq:redUnits}) for the case of gold. From top to bottom, the panels show the RDFs for state points that are respectively isomorphic, isothermal, and isochoric to the initial state points. The top panel shows the data collapse along an isomorph as predicted by isomorph theory. Isomorph and isotherm share the state point indicated by the black line, while isochore and isomorph match at the purple line (this is done to avoid melting of the crystal along the isochore).}
\label{rdf}
\end{figure}

\begin{figure}
\centering
    \includegraphics[width=8.66cm]{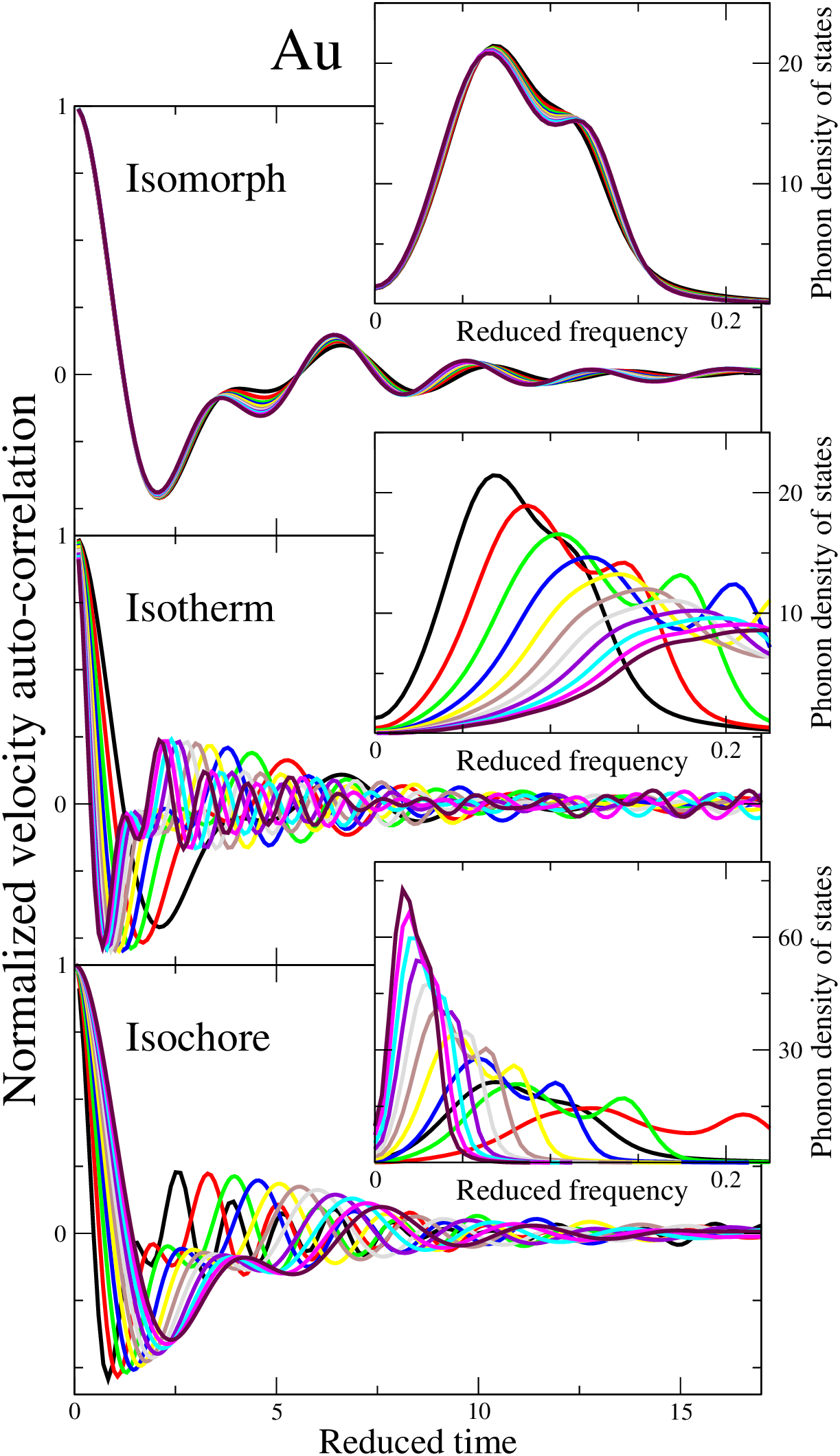}
\caption{The figure shows the normalized velocity autocorrelation function (VAF) 
for the same state points as in Fig.~\ref{rdf}. The insets depict the respective phonon densities of states obtained from the Fourier transform of the VAF (Eq.~\eqref{eq:dos}). The predicted data collapse along the isomorph is still obvious, although not as good as in the RDF case.}
\label{vaf}
\end{figure}

\newcommand{\moveup}{\vspace{-7.2mm}}
\begin{figure*}[t]
\begin{minipage}{.445\textwidth}

    \includegraphics[width=\linewidth]{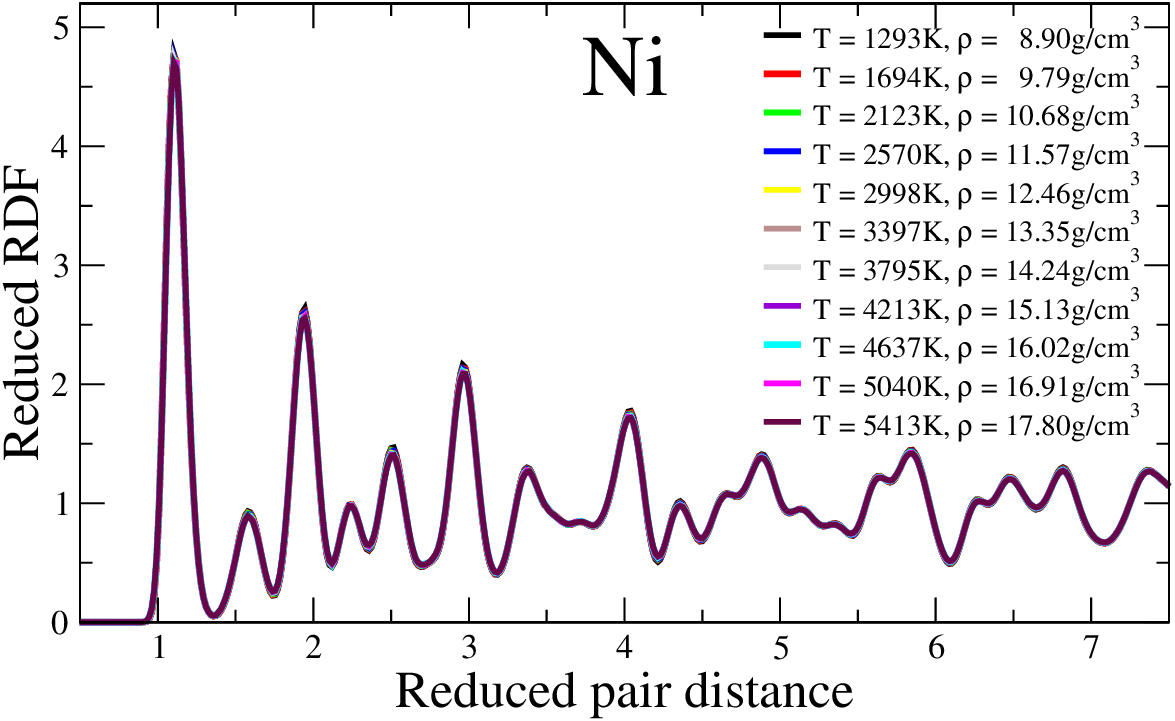}\moveup

    \includegraphics[width=\linewidth]{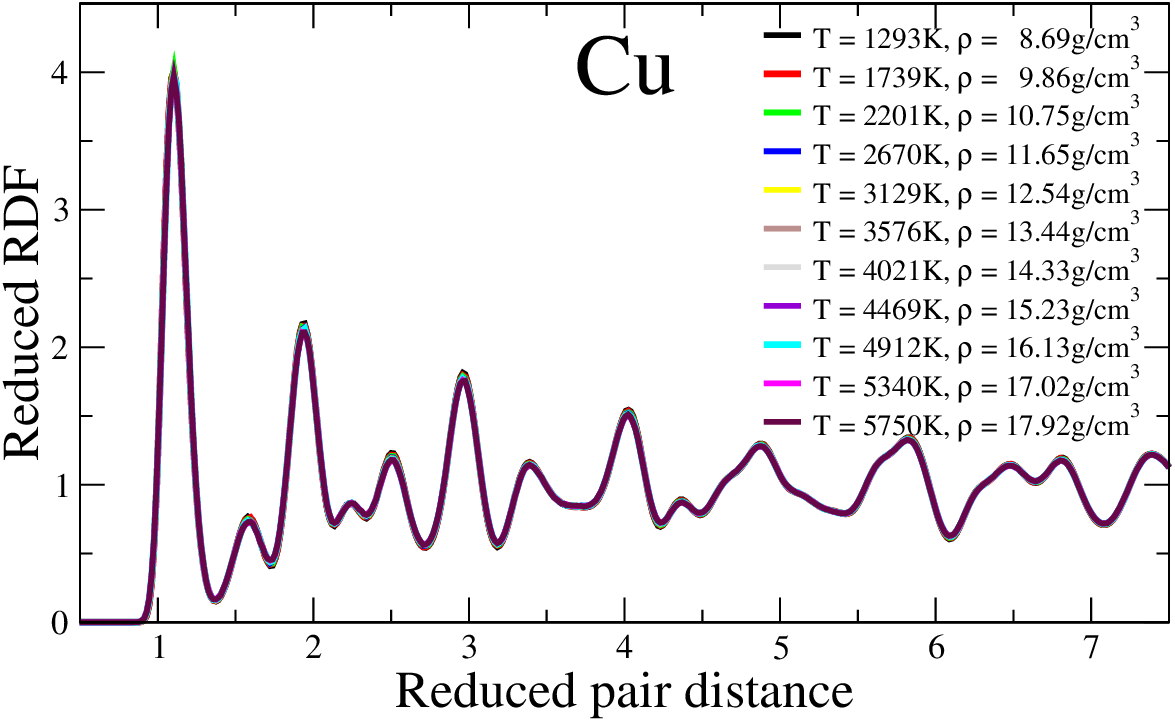}\moveup

    \includegraphics[width=\linewidth]{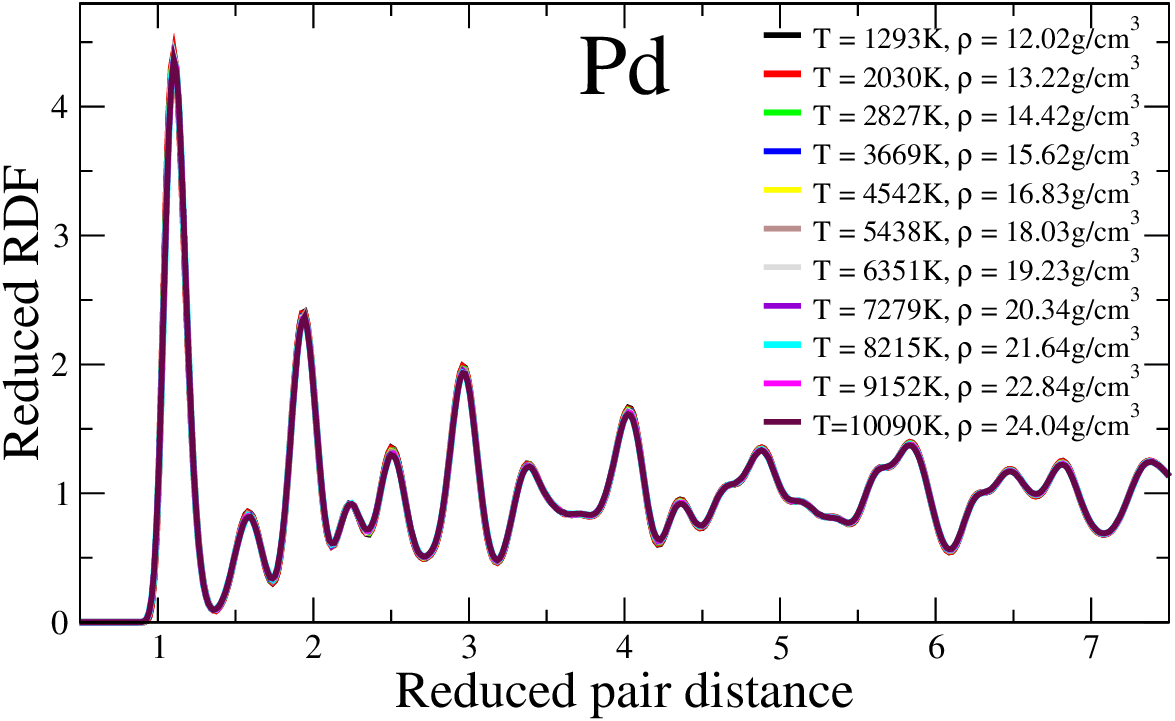}\moveup

    \includegraphics[width=\linewidth]{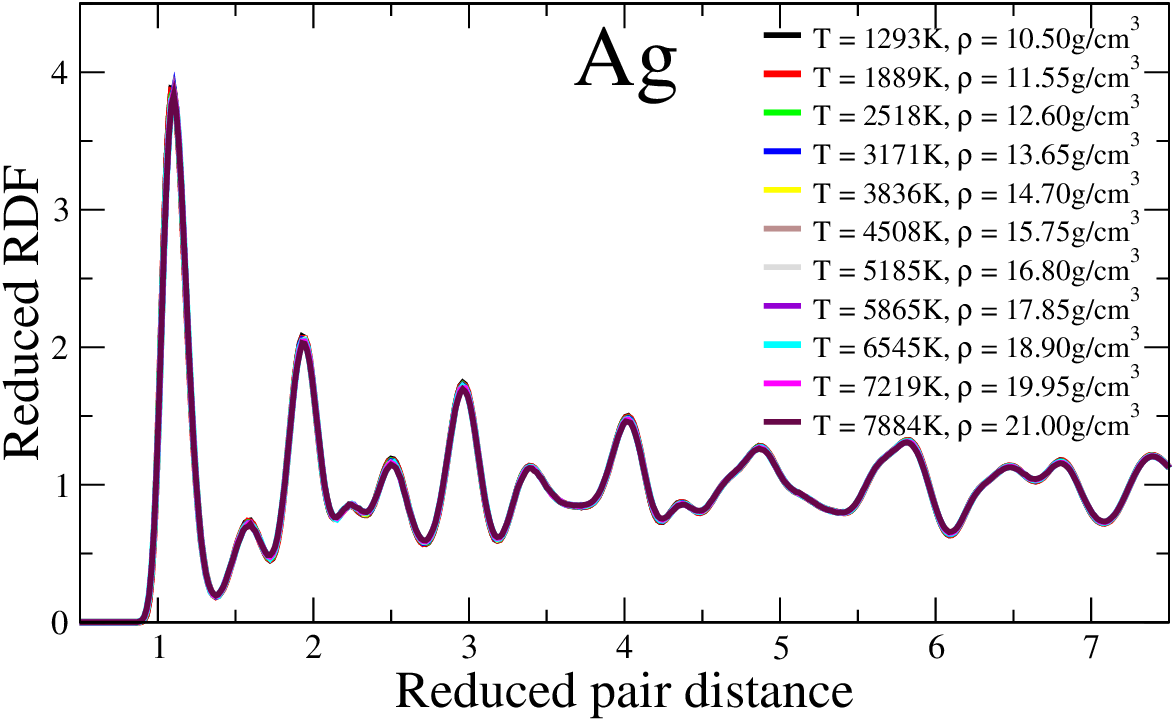}\moveup

    \includegraphics[width=\linewidth]{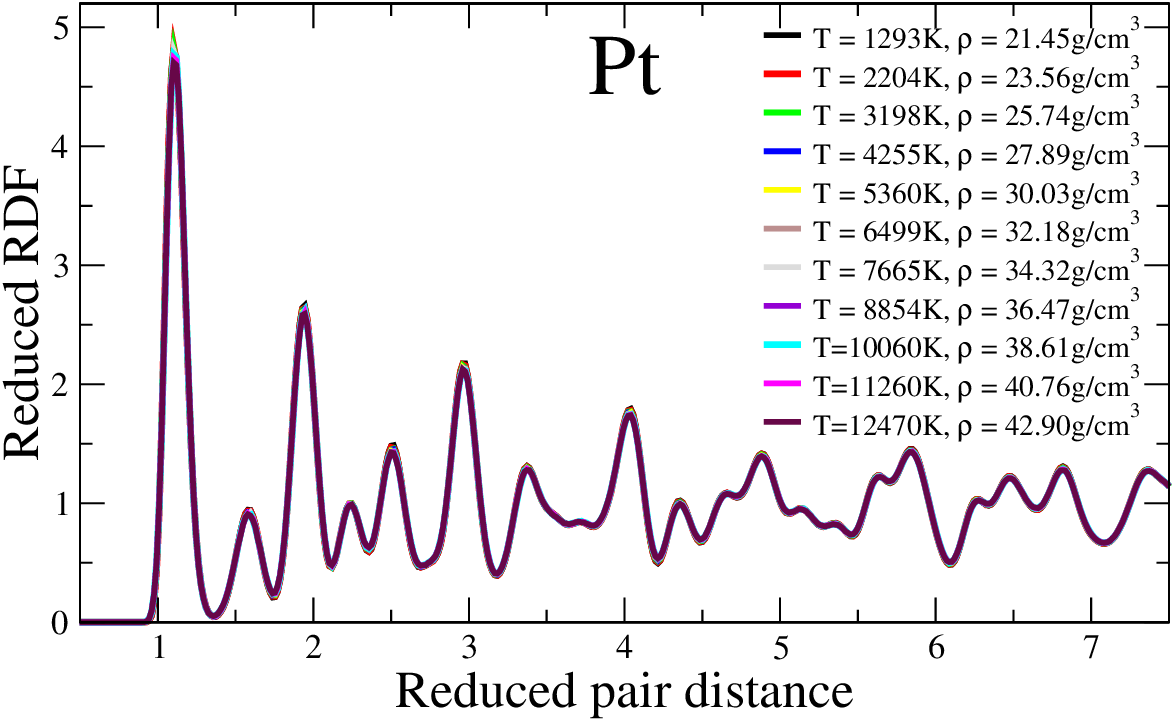}

\end{minipage}\hspace{17pt}
\begin{minipage}{.46\textwidth}

    \includegraphics[width=\linewidth]{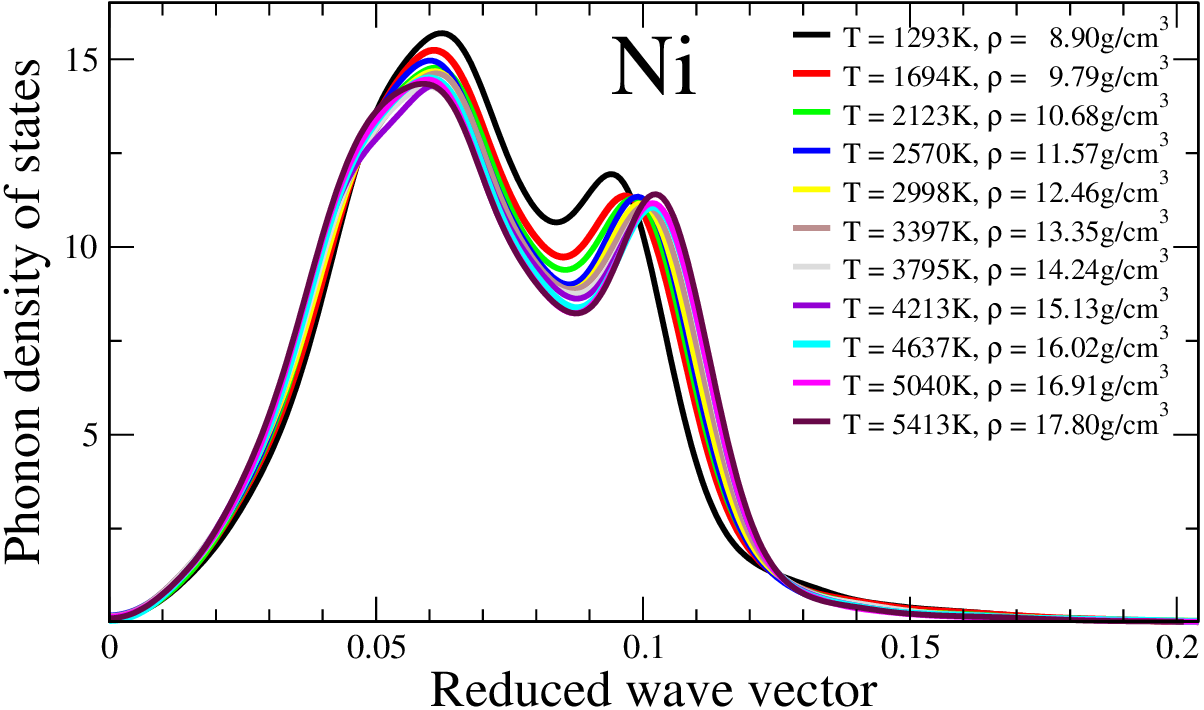}\moveup

    \includegraphics[width=\linewidth]{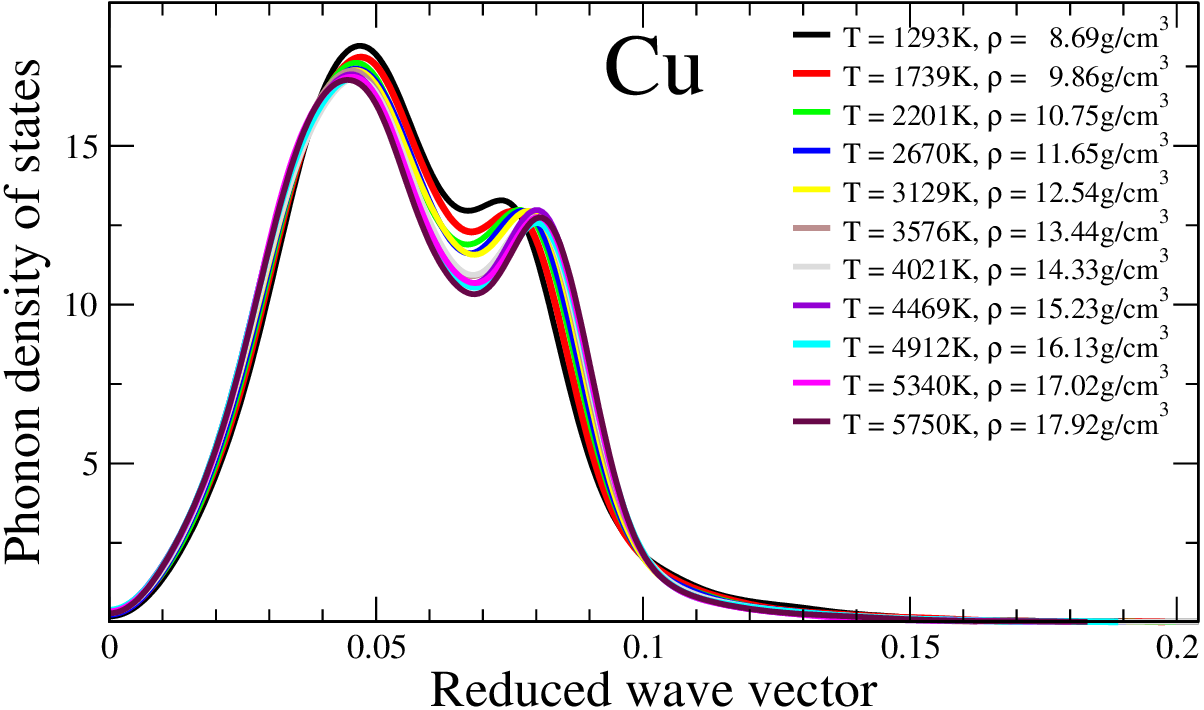}\moveup

    \includegraphics[width=\linewidth]{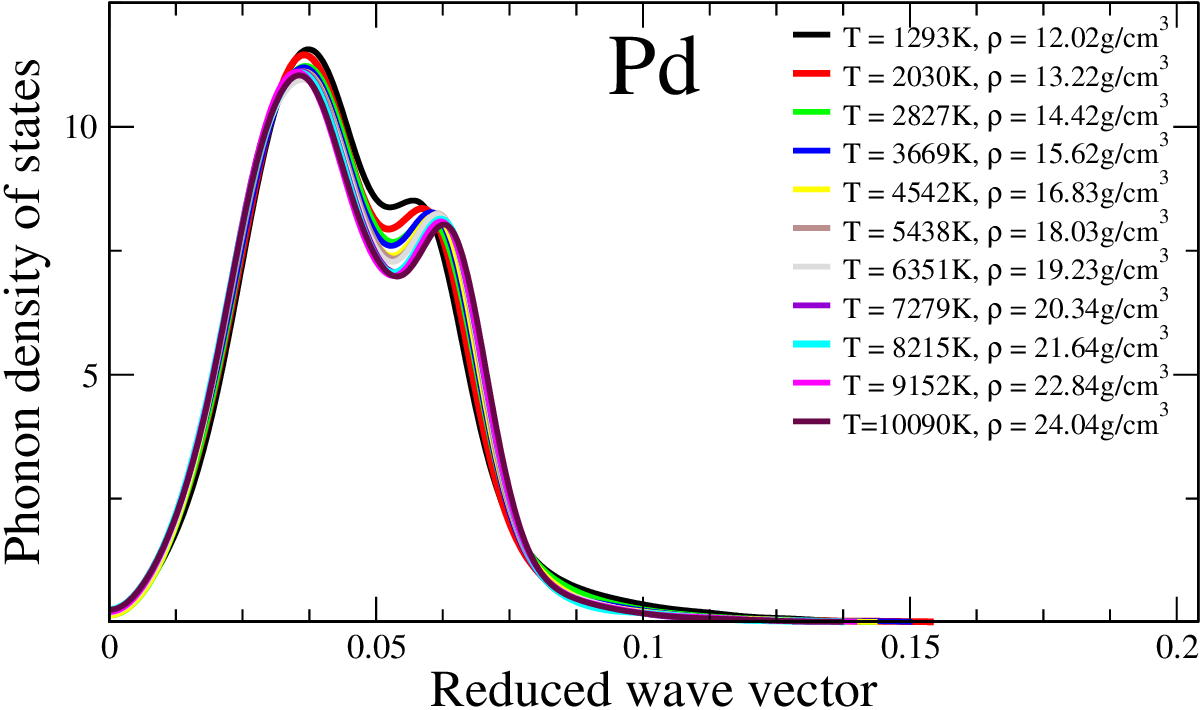}\moveup

    \includegraphics[width=\linewidth]{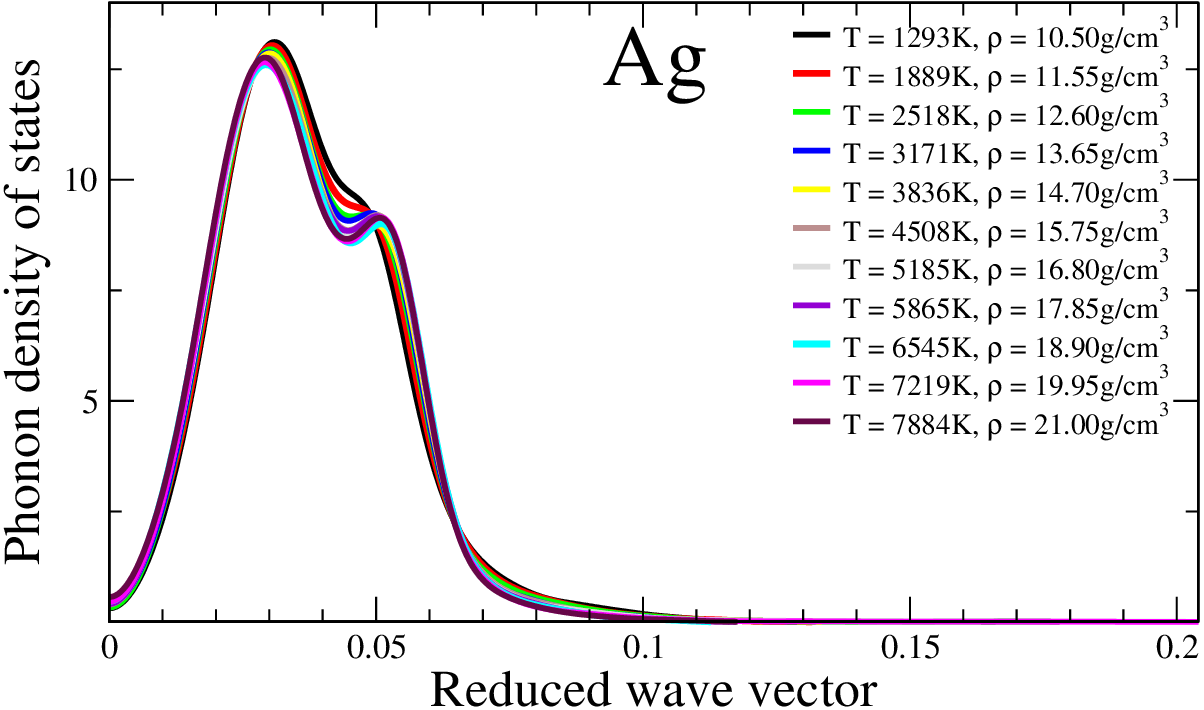}\moveup

    \includegraphics[width=\linewidth]{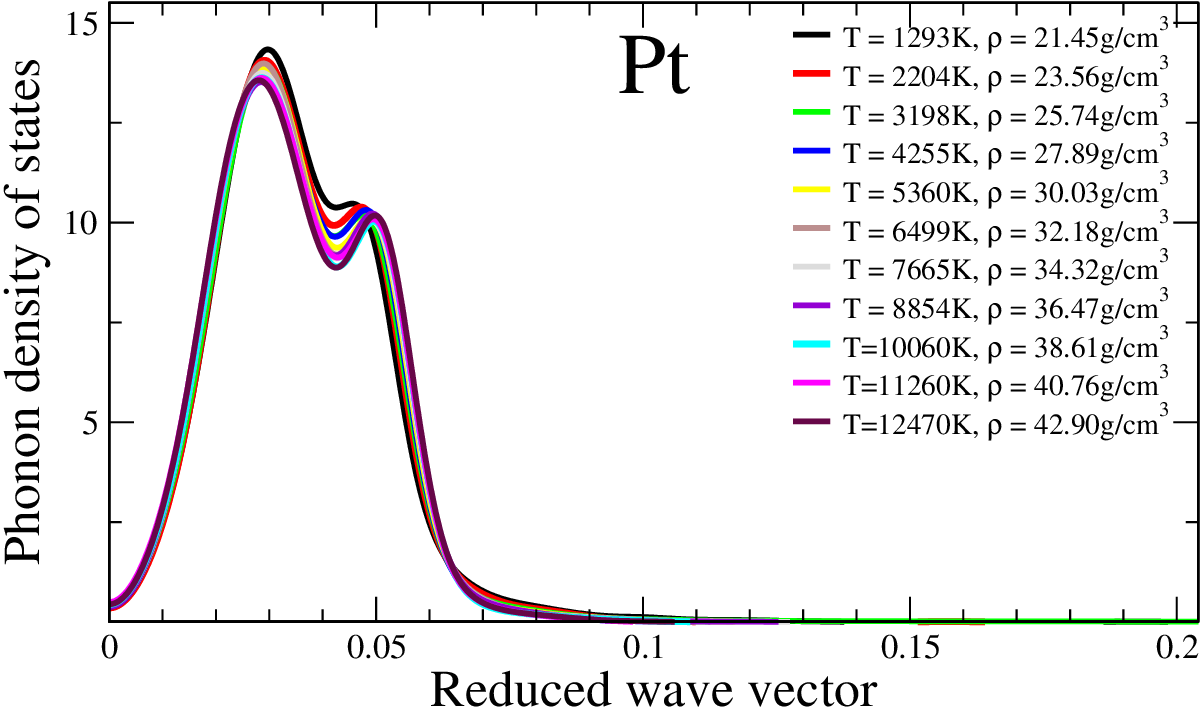}
    
\end{minipage}
\caption{Radial distribution functions and phonon densities of states along isomorphs for Ni, Cu, Pd, Ag, Pt. The other five metals confirm the findings for gold, i.e., a near perfect collapse for the structure and a less perfect, but reasonable collapse for the dynamic. The most notable deviations are for the first state point(s).}
\label{allMats}
\end{figure*}

We start with the results on structure and dynamics. For brevity, the results shown here in detail are from simulations for gold; the other five materials exhibit the same behaviour and will be presented in a summarized fashion. A phase diagram is shown in Fig.~\ref{Phase-Diag} indicating the isomorph simulated along with the melting curve for the model, to give an idea of where in the phase diagram our focus lies. Some numerical data for gold along the isomorph are shown in Table.~\ref{tab:Au_data}. 
The structure of a system can be quantified by the radial distribution function (RDF), also called pair-correlation function $g(r)$, which is a measure of the probability of finding a particle at a distance $r$ away from a given reference particle. Figure \ref{rdf} shows the RDF for the reduced pair distance $\tilde{r} = \rho^{1/3} r$ for the state points indicated in the panels, thus along an isomorph, an isotherm, and an isochore respectively. The peak's positions are expected to remain the same also along the isotherm and isochore as a trivial consequence of the reduced pair distance being scaled by $\rho^{1/3}$. Isomorph theory predicts that the structure along an isomorph is invariant, thus we expect all isomorphic RDF's to collapse onto a single curve. Figure \ref{rdf} validates this with very good approximation, even for large density changes, for the case of gold.

In addition to the structure, also the dynamics of isomorphic state points are predicted to be invariant. The dynamics are studied here by means of the velocity autocorrelation function (VAF). Figure \ref{vaf} shows the normalized reduced-unit single-particle VAFs obtained from the same simulations and state points of gold as the RDF data. The top, middle and bottom panel show the VAFs for state points along isomorph, isotherm and isochore, respectively. %
The isomorphic curves exhibit a reasonable collapse, but with some deviation especially compared to the near perfect agreement in the RDF case. 
The insets of Fig.~\ref{vaf} show the phonon (vibrational) density of states of their respective curves. The spectrum is related to the Fourier transform of the velocity auto-correlation function via \cite{Changyol/others:1993}:

\begin{equation}
    \rho (\omega) = \frac{1}{3N T k_B} \int_{-\infty}^{\infty} \sum_{i=1}^N \langle v_i(t) v_i(0) \rangle C(t) \exp^{i \omega t} dt
    \label{eq:dos}
\end{equation}
where we include a Gaussian function $C(t)=\exp(-(t/t_c)^2)$ (with $t_c$ invariant in reduced units) to smoothly truncate the integrand, which otherwise decays very slowly compared to the data-sampling window. 


We obtained similar results for structure and dynamics for the five other materials simulated - viz., Ni, Cu, Pd, Ag and Py, see Fig.~\ref{allMats}. Each row shows the RDF on the left and the phonon density on the right, along an isomorph for one metal. All metals demonstrate a comparably good collapse to that found for gold.

\subsection{Other implications of isomorph theory}


\begin{figure}
\centering
    \includegraphics[width=8.66cm]{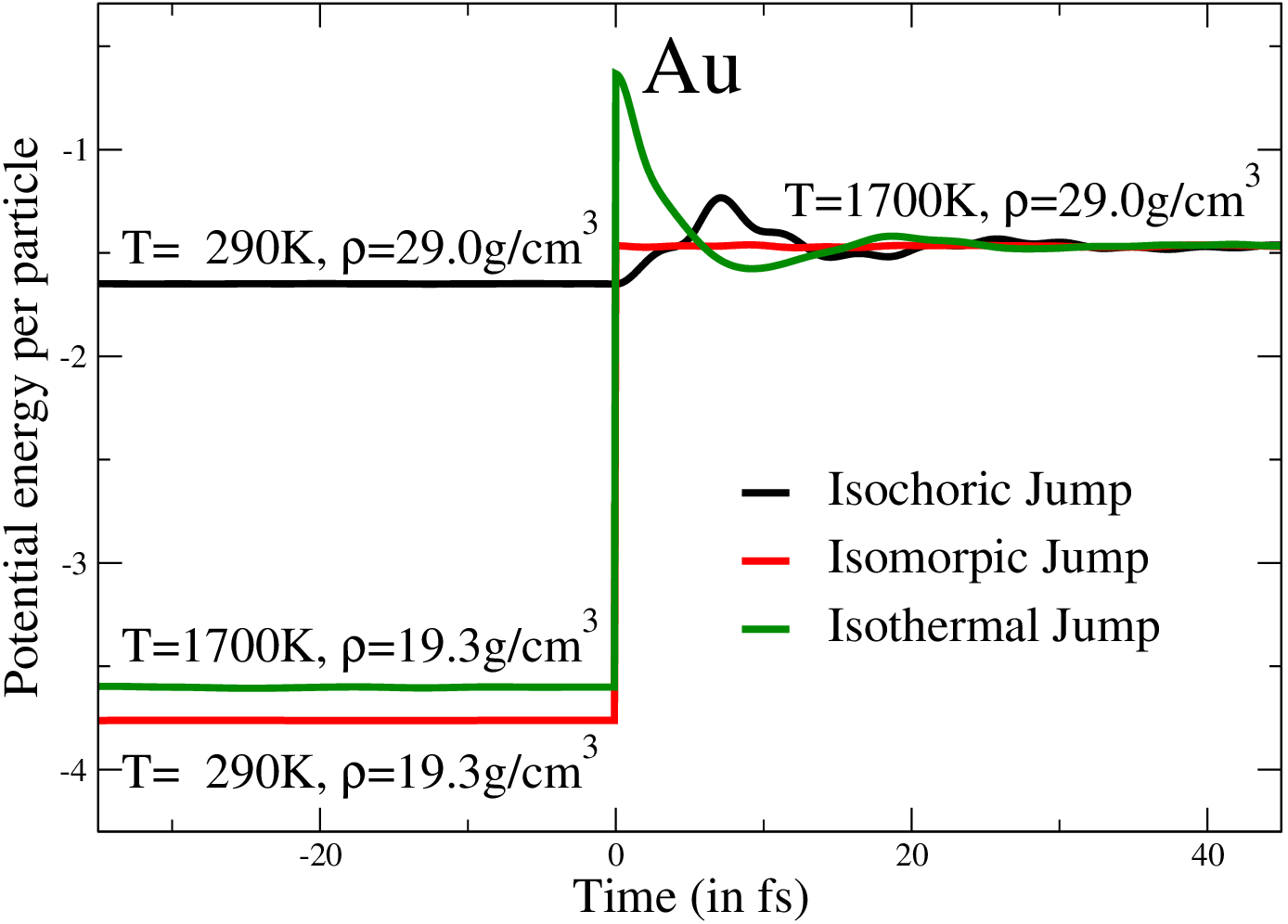}
\caption{Potential energy per particle before and after instantaneous jumps at $t=0$ between the state points indicated in the figure. The black and the green line depict jumps between points that are, respectively, isochoric and isothermal to each other. Only the red line shows instantaneous equilibration after the jump, as expected for isomorphic state points.}
\label{jump}
\end{figure}

Another prediction from isomorph theory concerns {\em isomorph jumps}, which refers to a sudden change in density via a uniform scaling of all article coordinates, and temperature between state points on the same isomorph. Isomorph theory implies that such a sudden change from a well equilibrated initial state point should not require further equilibration post jump, predicting the system to be instantaneously in equilibrium at the final state point ~\cite{paper4}, because the Boltzmann probabilities are unchanged by a jump along an isomorph. This prediction has been validated for viscous liquids, as well as perfect Lennard-Jones crystals. There is, however a subtle difference between these two cases.


The simulations start with runs at the respective starting point to make sure the systems are in equilibrium. %
%
At the `jump', the density is changed by uniformly scaling all particle coordinates, the temperature for the thermostat is set to the new value and all velocities are scaled accordingly. The results for gold can be found in Figure \ref{jump}, showing the potential energy per particle before and after jumps to the point indicated in the figure. The initial points have been chosen to be isothermal (green), isochoric (black) and isomorphic (red) to the final state point.
The red line clearly validates the prediction as it shows no changes in potential energy post jump, thus the system is in equilibration right away. In contrast to this, the black and the green lines are clearly not in equilibrium and the potential energies oscillate towards the new level.


\begin{figure}
\centering
    \includegraphics[width=8.66cm]{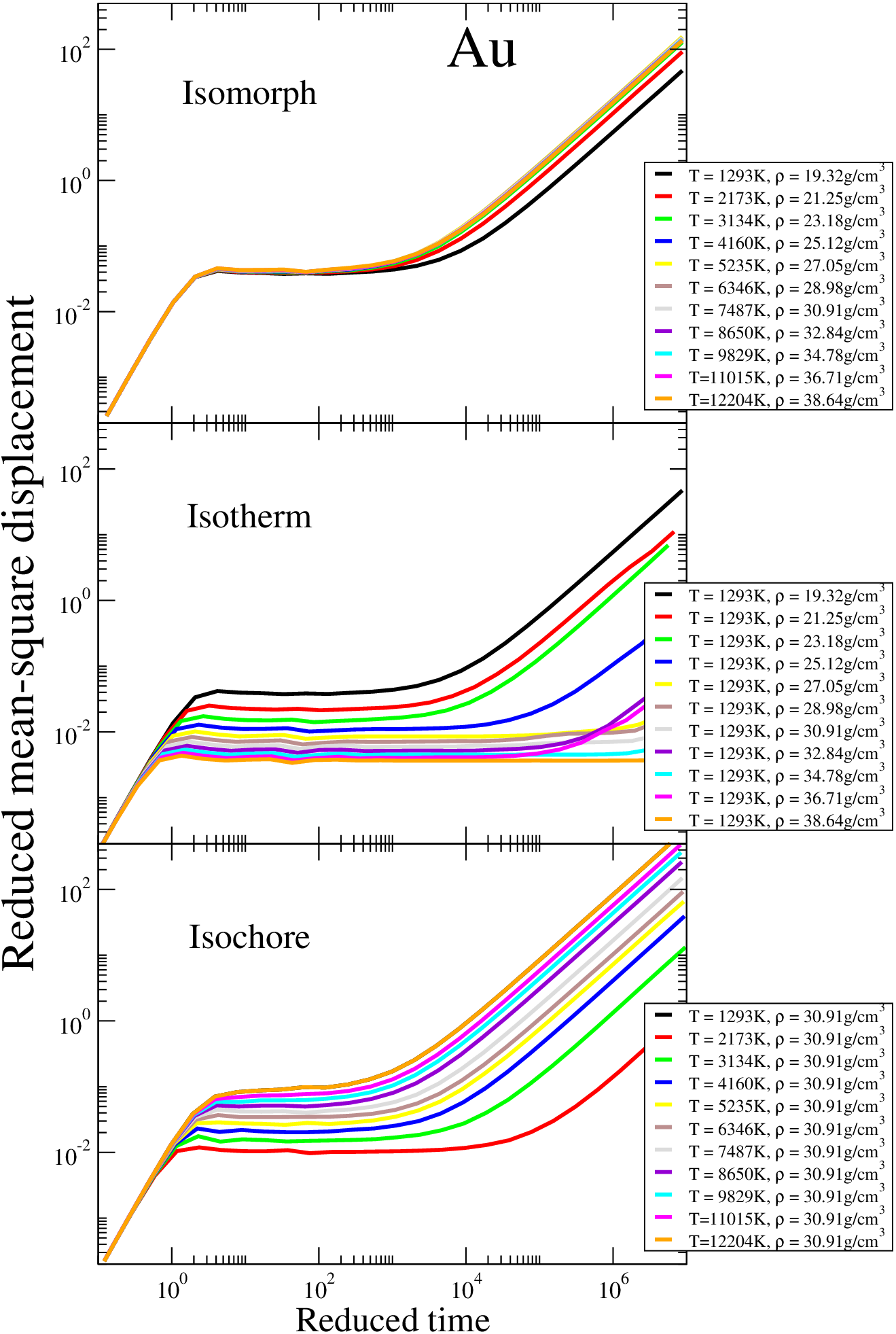}
\caption{Mean-squared displacement (MSD) for the state points indicated. To obtain vacancies, four randomly selected atoms have been removed from the initial crystal. The collapse exhibited along the isomorph is good in the ballistic regime (trivial) and the plateau (less trivial) while the diffusive part shows some deviations, especially for the lowest density/temperature state point. 
}
\label{vacancies}
\end{figure}

\begin{figure}
\centering
    \includegraphics[width=8.66cm]{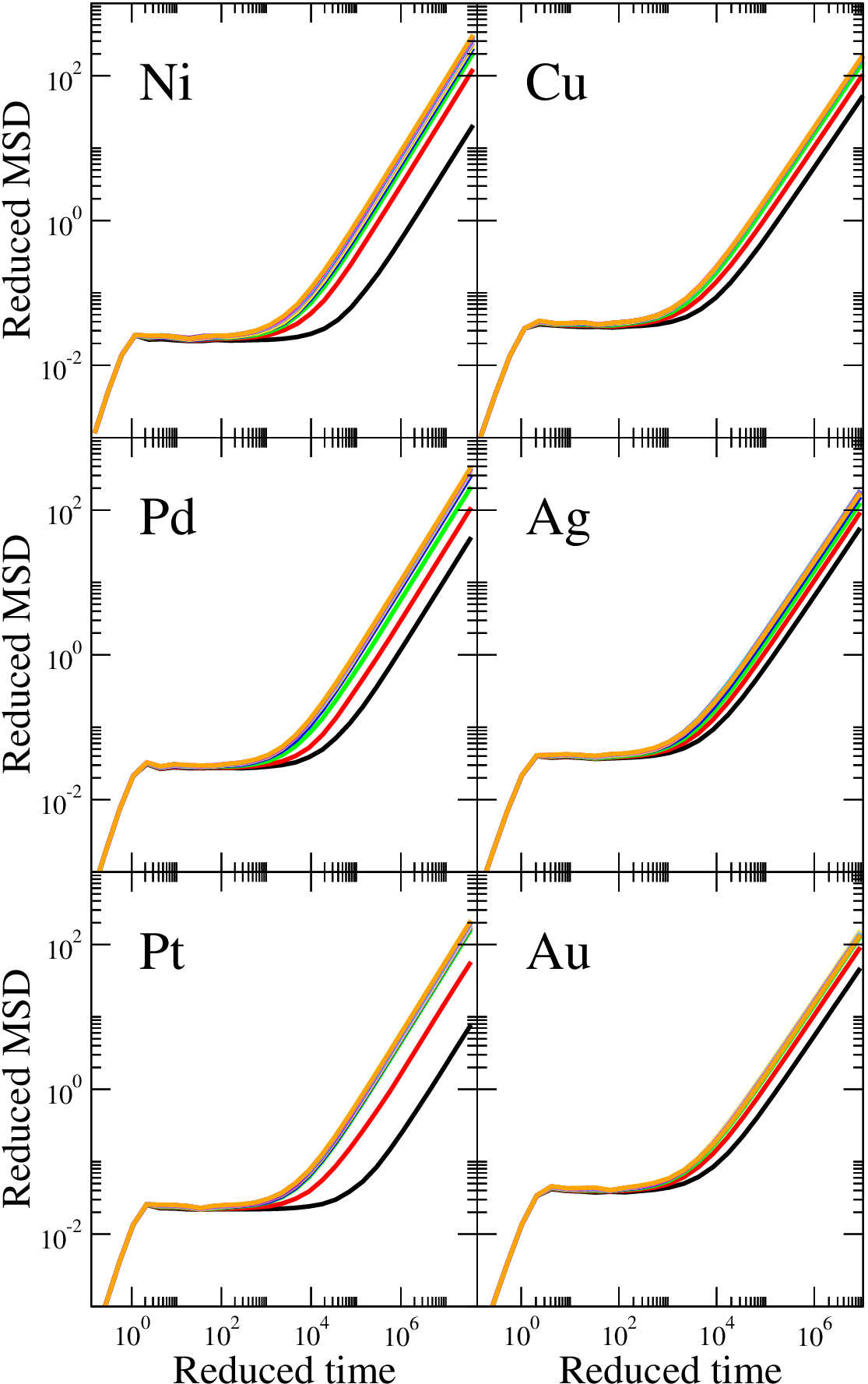}
\caption{Mean-squared displacement along isomorphs for the six metals (the bottom right panel depicting gold shows the same data as the top panel of Fig.~\ref{vacancies}).
}
\label{all_vacancies}
\end{figure}


Since many mechanical properties in crystals are associated with the existence of defects in the lattice and these properties are also expected to be isomorph invariant, we will examine this in the following for the case of vacancy diffusion.
A vacancy in the lattice is an empty spot from which the atom has been removed. This introduces a new kind of dynamics to the crystal since atoms can now jump to the new, empty positions on the lattice, resulting in the vacancies moving around. 
%
Vacancy diffusion is quantified by means of the mean-square displacement (MSD) of the atoms\cite{Albrechtsen2014, Vaari:2015}. Figure \ref{vacancies} shows the MSD along isomorph, isochore and isotherm for the case of four particles removed from a $10\times 10\times 10$ fcc crystal of gold, corresponding to a vacancy concentration of 10$^{-3}$, which is only slightly higher than the experimental concentration close to melting, $7 \times 10^{-4}$ \cite{Simmons:1962}.

The figure shows an approximate collapse along the isomorph---although there is a noticeable deviation for the first two curves (the lowest two densities). The collapse is poorer than that seen for the Lennard-Jones crystal in Ref.~\onlinecite{Albrechtsen2014}. This can be partly explained by observing that the starting state point in the present case is closer (in density) to the triple point than was the case for the Lennard-Jones results, although the pressure in our simulations is large by experimental standards (see Table~\ref{tab:Au_data}). 
It is interesting to note that the pre-diffusive parts of the curves collapse much better. For the initial ballistic regime the MSD is proportional to time squared, which is a trivial consequence of the use of reduced units and has nothing to do with isomorph invariance (it is seen also for the isochore and isotherm). But the invariance of the height and the location of the onset of the plateau are non-trivial aspects of the vibrational dynamics. The diffusivity (corresponding in the double-logarithmic representation to the height of the long-time part of the MSD curves) is presumably determined by a single energy barrier associated with vacancy hopping. The poor collapse of the curves here therefore implies that this energy barrier scales in a slightly different manner than the potential energy surface near the ground state---it is the latter which controls vibrational dynamics whose fluctuations were used to determine the isomorph.

Figure~\ref{all_vacancies} shows the MSD along an isomorph in all six fcc metals with each crystal having four vacancies. The bottom right panel shows the same gold isomorph as in the previous figure. The same overall behaviour can be observed in the other metals as well, i.e., that the higher density/temperature points collapse well while the first (two) curve(s) exhibit an outlier behaviour. This is more notable for the materials on the left hand side; these have incomplete d-shells, corresponding to stronger bonding and higher melting points, therefore the simulated isomorphs (which all start at the same temperature) are further below the melting line in these cases.

Results for one and 16 vacancies in the same $10\times 10\times 10$ fcc crystal of the six metals can be found in the supplemental material. Both cases exhibit a much worse collapse than the case of four vacancies. For 16 vacancies, visualization (Snapshot included in the supplement material) shows that the vacancy concentration is to high which causes them to cluster together early on in the simulation runs. We thus inadvertently probed void migration rather than vacancy diffusion. Especially interesting is the case of one vacancy where clustering is not an issue. We found a failure to collapse much like Albrechtsen and Olsen found for LJ crystals with only one vacancy  \cite{Albrechtsen/Olsen:2013}. This case seems especially sensitive to departures from isomorph invariance (see also the discussion of Fig.~\ref{configuration-specific-gamma}).



\section{Discussion}

The overall results presented here are consistent with expectations from the work of Hummel {\it et al.}, which showed that most metals in the liquid state have a high virial potential-energy correlation coefficient $R$ (Eq.~\eqref{eq:R}) and are therefore R-simple. As such they are expected to have good isomorphs. The present work has concentrated on the crystal phase at moderate and high temperatures to avoid quantum effects. The analysis is similar to that undertaken by Albrechtsen {\em et al.} for Lennard-Jones and other simple model systems including simple molecules \cite{Ingebrigtsen/Schroeder/Dyre:2012}.
 
The basic predictions of isomorph theory are invariance of structure and dynamics when the observables are expressed in reduced units: lengths in terms of the interparticle spacing $\rho^{-1/3}$, energies in terms of the temperature $k_B T$, and times in terms of the time a particle with the thermal velocity would take to move an interparticle spacing (Eq.~\eqref{eq:redUnits}). With these units we find an excellent collapse of the radial distribution function. For the dynamics of the perfect crystal we studied the velocity autocorrelation function (VAF) and its Fourier transform, which can be interpreted as an effective vibrational density of states (VDOS). Here we observed an approximate collapse, clearly worse than the RDF, and also worse than the collapse seen for the Lennard-Jones crystal in Ref.~\onlinecite{Albrechtsen2014}. We validated the prediction of instantaneous equilibration for isomorph jumps. To study dynamics beyond vibrations we simulated a system with vacancies and monitored the mean squared displacement. The collapse here was also approximate, in fact poorer than for the VAF, suggesting that the relevant energies (around the saddle point of the vacancy hopping process) scale somewhat differently with density than energies near ground state (perfect crystal) which are relevant for vibrations. In particular one can imagine that the local density experienced by the hopping atom at the top of the energy barrier is quite different from the densities of the surrounding atoms, corresponding to different effective $\gamma$.

\begin{figure}
    \centering
    \includegraphics[width=8.66cm]{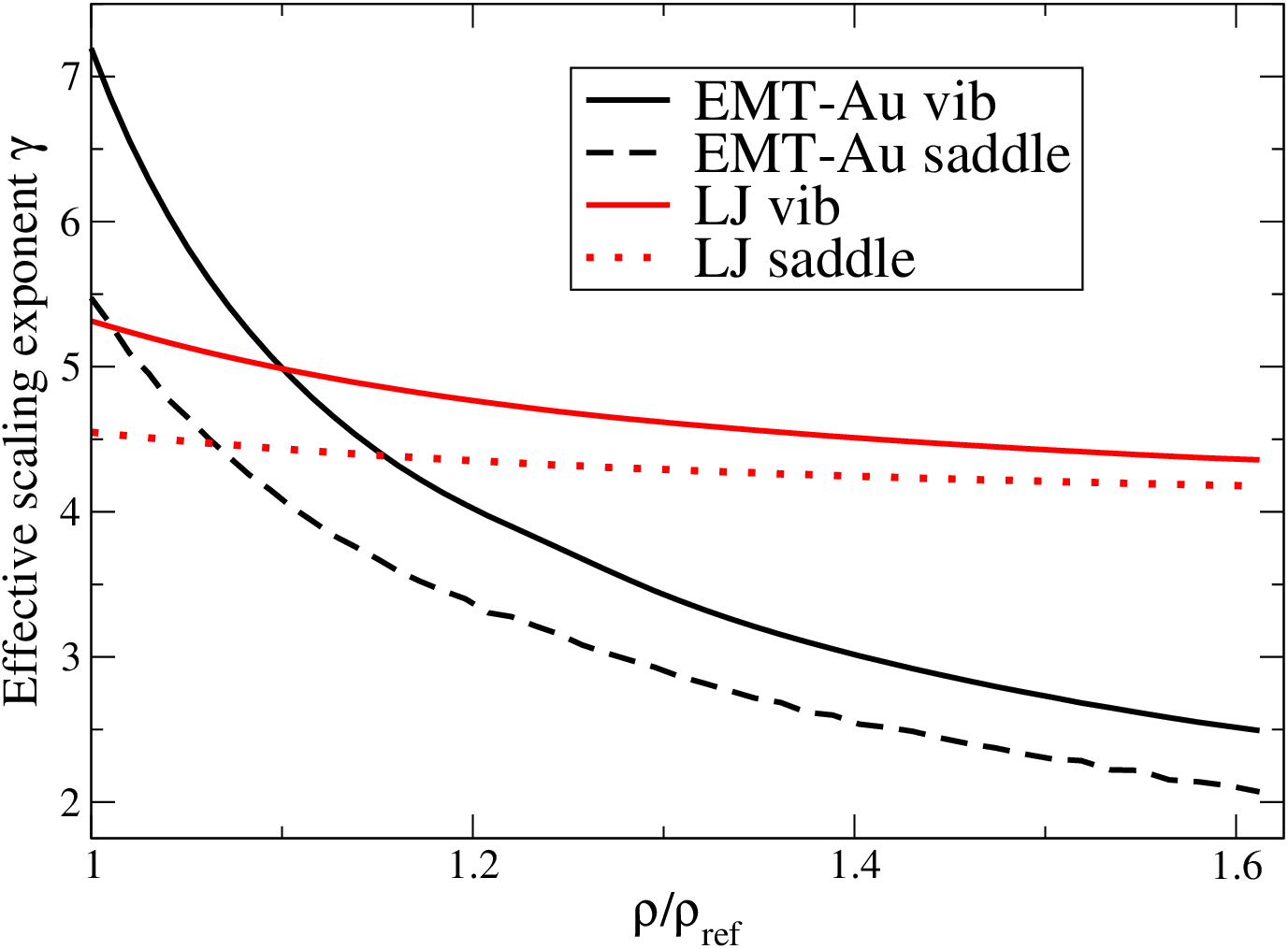}
    \caption{Effective scaling exponent for specific configurations determined using EMT potential for gold and the Lennard-Jones (LJ) potential. The vibrational configurations, one for each system, were sampled from an NVT run of the perfect crystal at density $\rho_{ref}=19.3$g/cm$^3$ and temperature 1300K for Au, density $\rho_{ref}=1.05\sigma^{-3}$ and temperature 0.630 $\epsilon/k_B$ for LJ. The potential energy relative to the perfect lattice was determined for a range of densities (scaling both the perfect lattice and the vibrational configuration). The saddle configurations are the unrelaxed saddle point between vacancy hopping, constructed by moving a neighbor atom of the vacancy in an otherwise perfect crystal halfway towards the vacant site. The energy difference between unrelaxed saddle point and the unrelaxed vacancy was determined for the same range of densities. The effective scaling exponents given by the logarithmic derivatives $d\ln E/d\ln \rho$ are plotted against the reduced densities $\rho/\rho_{ref}$.}
    \label{configuration-specific-gamma}
\end{figure}

In particular, the combination of locally high density at the saddle point for the hopping atom and the strong density dependence of the scaling exponent for EMT systems suggests a scenario like this. As a crude test of this we consider the energy of the ``unrelaxed saddle point'' relative to that of the unrelaxed vacancy, as well as the energy of a typical vibrational configuration of the defect-free lattice, drawn from a simulation at a specific temperature, relative to that of the perfect lattice. The unrelaxed vacancy is the perfect lattice with one atom removed. The unrelaxed saddle point is the configuration obtained by displacing a neighbor of the removed atom exactly halfway towards the empty site. The logarithmic derivatives of these energy differences give a kind of ``configuration-specific'' scaling exponent $\gamma$, plotted in  Fig.~\ref{configuration-specific-gamma}. There is a significant reduction in the effective scaling exponent for the unrelaxed saddle-point energy compared to that of the vibrational energy (5.5 versus 7.2 at the initial density). Since vibrational fluctuations dominate the determination of the $\gamma$ used to generate isomorphs, the lower scaling exponent for saddle points means these configurations have a lower energy than expected as one moves along the isomorph, which is why the mean squared displacement in reduced units is higher than than for the reference state point. The contrast between the energy fluctuations used to determine the isomorph and those relevant for the dynamics, is greater for a defective crystal than for a liquid or amorphous solid, allowing such deviations from perfect isomorph scaling to arise. The figure includes also results of the same calculation for the Lennard-Jones crystal, where there is also a difference, albeit smaller than in the EMT case. The presence of this difference is consistent with the lack of collapse for a single vacancy noted in Ref.~\onlinecite{Albrechtsen/Olsen:2013}, while its small size explains the generally better collapse found in Ref.~\onlinecite{Albrechtsen2014}. It remains somewhat unclear why including several vacancies then gives a better collapse; it presumably involves the interactions between them (including vacancy binding/unbinding) reducing the contrast between the energy fluctuations used to determine the isomorph and the relevant saddle-point energy which governs vacancy dynamics. In a sense it is not that surprising that the specific parts of the potential energy function associated with vacancy hopping behave differently under density changes compared to those related to vibrations. Thus, while less pronounced in the Lennard-Jones case, the same deviations occur in both systems.

%
%
The general degree of isomorph invariance is similar for the different metals (see Fig.~\ref{allMats} for the radial distribution functions and phonon density of states, and Fig.~\ref{all_vacancies} for the vacancy diffusion), which is not surprising since the same functional form of interatomic interactions is used for all of them. In the future, it is important to investigate isomorph invariance of these metals using other types of potentials for example EAM.

\begin{figure}[t]
    \includegraphics[width=8.66cm]{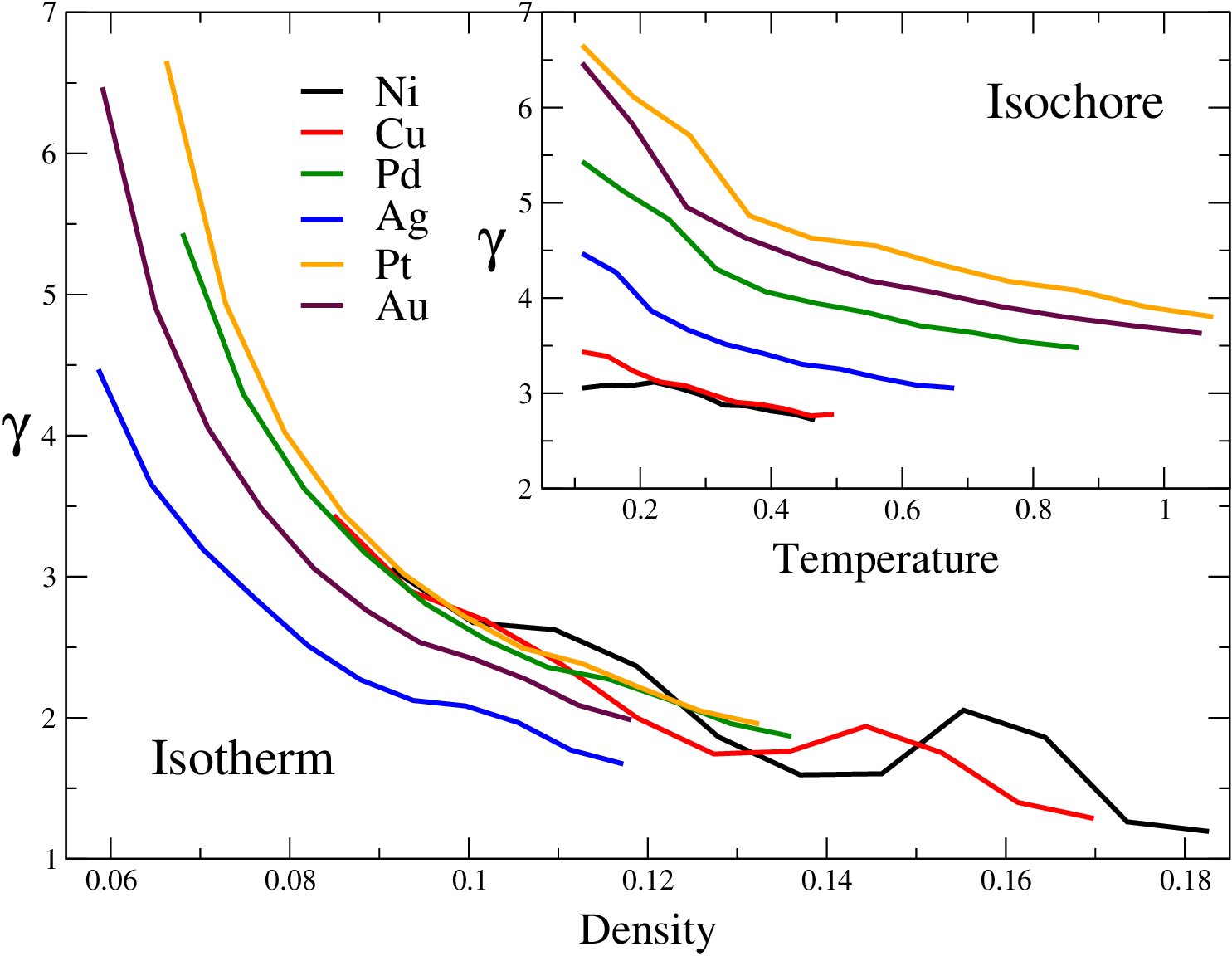}
\caption{Variation of $\gamma$ for the six fcc metals along an isotherm in the main panel and along an isochore in the inset. The $\gamma$ variations are clearly dominated by changes in density. The bumps visible in the low $\gamma$ regime, most notable in Cu and Ni, are due to the cut-off as detailed in the supplemental material.}
\label{T/roh_gamma}
\end{figure}

Unlike in other systems, the scaling exponent $\gamma$ is strongly state point dependent when using the EMT potential. The main panel in Figure \ref{T/roh_gamma} shows the variation of $\gamma$ for the six fcc metals with changing density at constant temperature. The behaviour for increasing temperature at fixed density can be seen in the inset. It is evident that the change in $\gamma$ is dominated by changing density and only mildly decreasing with temperature. Thus, the variation of $\gamma$ along an isomorph (not pictured) displays a similar behaviour to that of the isotherms. The oscillatory behaviour along the isotherms in the low $\gamma$ region, most obvious in the cases of Ni and Cu, is an artifact due the cutoff and occurs when increasing the density pushes a new neighbour shell through the cutoff distance (see supplement).

\begin{figure}[t]
\includegraphics[width=8.66cm]{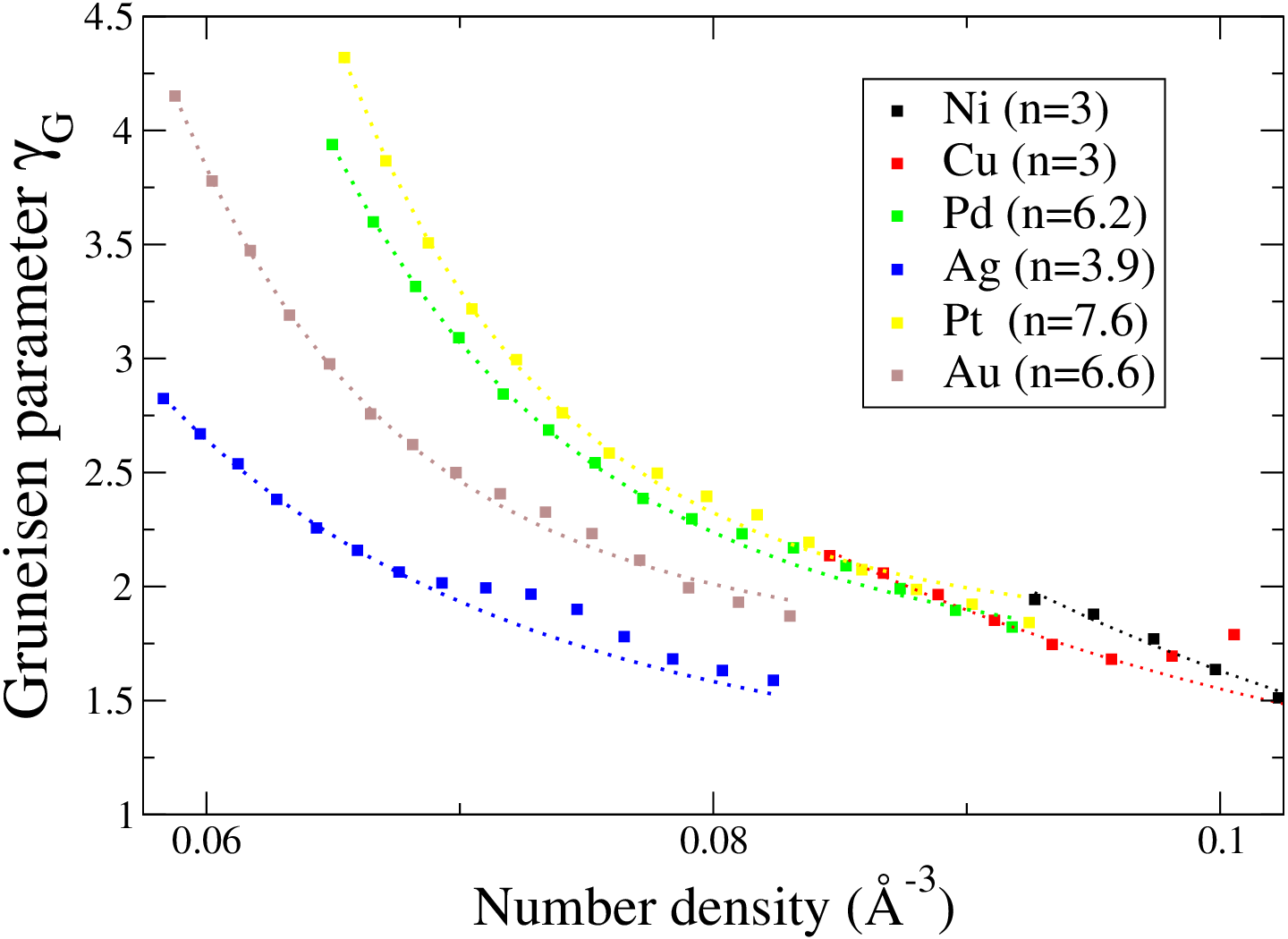}
\caption{\label{gruneisen_fits}Gruneisen parameter $\gamma_G$ for the six metals as a function of (number) density at temperature 300 K (squares). The dotted lines show fits to the functional form $a+b/\rho^n$, with the exponent $n$ indicated for each metal in the legend. The data show small bumps associated with the same cutoff artifact mentioned before; for the fitting only the data at small densities was used. For Cu and Ni allowing the exponent to vary leads to negative values of the additive constant $a$ and so here the exponent was fixed at $n=3$.}
\end{figure}

\begin{table}[b]
    \caption{\label{gruneisen_comparison} Comparison of experimental Gr\"uneisen parameters with EMT-values at ambient pressure and temperature.}
    \addtolength{\tabcolsep}{5pt}
    \vspace{5pt}
      \begin{tabular}{|c|c|c|}
        \hline
        ~Sym~ & ~$ \gamma_G$ (exp.)~ & $ \gamma_G$ (EMT) \\
        \hline
        Ni & 1.88 \textsuperscript{(a)} & 1.9 \\
        Cu & 1.96 \textsuperscript{(a)} & 2.1 \\
           & 1.99 \textsuperscript{(b)} &     \\
        Pd & 2.33 \textsuperscript{(c)} & 3.4 \\
        Ag & 2.40 \textsuperscript{(a)} & 2.8 \\
           & 2.33 \textsuperscript{(b)} &     \\
        Au & 2.94 \textsuperscript{(b)} & 4.2 \\
        Pt & 2.54 \textsuperscript{(a)} & 4.3 \\  
        \hline
      \end{tabular}
     \hspace{3pt}
     \begin{minipage}[b]{1.55cm}
     \flushleft
      \textsuperscript{(a)}~Ref.~\onlinecite{Kittel:1956},\\ %
      \textsuperscript{(b)}~Ref.~\onlinecite{Quareni/Mulargia:1989},\\ %
      \textsuperscript{(c)}~Ref.~\onlinecite{Smith/White:1977}
      \vspace{2em}
     \end{minipage}%


\end{table}

Next we discuss briefly the connection between the density scaling exponent  $\gamma$, of the two most important parameters (together with $R$) in isomorph theory and the Gr\"uneisen parameter $\gamma_G$, which is an important thermodynamic parameter in the study of solids. The latter plays a fundamental role in the Mie-Gr\"uneisen equation of state, often used to model metals at high pressures. Pandya {\it et al.} \cite{Pandya/others:2002} argue that the Gr\"uneisen parameter, involving as it does third derivatives of the potential, is a stringent test of a model of a solid. Ref.~\onlinecite{Hieu/Ha:2013} discusses the use of a pressure-dependent Gr\"uneisen parameter to estimate the melting curves of silver, gold, and copper at high pressure. Insight from isomorph theory and the study of the isomorphic properties of metals can help to understand the density dependence of $\gamma$ and by extension $\gamma_G$. The microscopic definition of $\gamma_G$ involves the density dependence of normal mode frequencies, but we focus on the macroscopic or thermodynamic definition

\begin{equation}
  \gamma_G \equiv V \frac{\alpha_p K_T}{C_V}
\end{equation}
where $\alpha_P$ is the thermal expansion coefficient, $K_T$ the isothermal bulk modulus, and $C_V$ the isochoric specific heat. The relation

\begin{equation} \label{gamma_gamma_G}
\gamma_G = \frac{\gamma C_V^{\textrm{ex}} + k_B}{C_V}
\end{equation}
between $\gamma$ and $\gamma_G$ - where $C_V^{\textrm{ex}}$ is the excess part of the isochoric heat capacity $C_V$ - was derived in \cite{paper3} and is exact within the classical approximation. Typically $\gamma$ is greater than $\gamma_G$ by around a factor of two. Using Eq.~\eqref{gamma_gamma_G},  Hummel {\it et al.} compared values of $\gamma$ determined from the experimental values of $\gamma_G$ for liquid metals to values determined from their DFT calculations (see their Figure 5). In Table~\ref{gruneisen_comparison} we compare values of $\gamma_G$ determine for the crystal phase at ambient temperature and pressure to values determined for EMT. We find good agreement for Cu and Ni, while the other values are significantly overestimated compared to experiment. From Table~\ref{gamma_R_DFT_EMT} for the comparison of $\gamma$ between EMT and DFT for the liquid, and from the work of Hummel {\em et al.} who compared DFT results for $\gamma$ with values inferred from experimental gruneisen parameters, we can infer that for Au and Ag the EMT values match the DFT values reasonably well, but both overestimate the experimental values of $\gamma$ and $\gamma_G$. For Pd and Pt the DFT results match experiment but the EMT results are too high. %

There is interest in the literature in the density-dependence of $\gamma_G$, for example for understanding the state of matter deep in the earth's interior \cite{Quareni/Mulargia:1989,Pandya/others:2002}. A frequently-used empirical model for the density dependence is $\gamma_G\rho=constant$, i.e. the Gr\"uneisen parameter decreases inversely with density. This is consistent with our observation that $\gamma$ is mainly a function of density and for EMT metals decreases strongly with density; however our data do not support a $1/\rho$ dependence of $\gamma_G$ (see Fig.~\ref{gruneisen_fits}). A closer look at the functional form of the EMT potential should provide some clues for the density dependence of both $\gamma$ and $\gamma_G$. 

Finally we discuss implications for the thermodynamics of melting and freezing of metals. An early prediction of the isomorph theory was that the melting curve follows an isomorph for R-simple systems \cite{paper4}. This follows from the general idea that the structure is invariant. Considering constant volume conditions in the coexistence region, ensuring the presence of a fixed amount of each phase, a broad interpretation of ``structure'' would include ``degree of crystallization'', and would have the consequence that the melting curve must follow an isomorph (otherwise the degree of crystallization along an isomorph would vary). However for realistic systems isomorph invariance applies to a single phase, but not a system containing two phases with different densities. In the latter case terms in the free energy which depend on density only become relevant, affecting the position of the melting curve while having no relevance for the structure and dynamics of a single phase. This has been studied in detail in Refs.~\onlinecite{Pedersen/others:2016a} and \onlinecite{Pedersen/others:2017a}. In particular the theory developed in Ref.~\onlinecite{Pedersen/others:2016a} allows calculation of the freezing and melting lines using isomorphs as the basis for a perturbative approach. Computer simulations confirmed the predictions for the Lennard-Jones case. The data in Fig.~\ref{Phase-Diag} for the melting curve seem not to coincide with crystal isomorph, though data for the freezing line for the same system (not shown) coincide very closely with a liquid isomorph. The methods of Ref.~\onlinecite{Pedersen/others:2016a} should allow both to be calculated from simulations at a single temperature. Applied to more computationally demanding first-principles methods, such as DFT this gives the potential to make accurate melting curve determinations at high pressures.\bigskip

In summary, we have shown that isomorph theory applies very well to fcc metals simulated using the effective medium theory many-body potential. We find the expected invariance of structure and, slightly less perfectly, of vibrational dynamics. The instanteous equilibration following an isomorph jump is also seen. Slightly larger deviations emerge when studying defect dynamics. This was argued to be a consequence of, on one hand, the contrast between the configurations governing (in this case) vacancy hopping and those dominating the fluctuations, and on the other hand, the strong density dependence of~$\gamma$.



%



\end{document}